\documentclass[nofootinbib,twocolumn]{revtex4-1}
\usepackage[utf8]{inputenc}
\usepackage{amsmath}
\usepackage{graphicx}
\usepackage{caption}
\usepackage{subcaption}
\usepackage{amssymb}
\usepackage{color}
\usepackage{cancel}
\usepackage{framed}
\usepackage{comment}
\usepackage{mathtools}
\usepackage{hyperref}
\usepackage{chngcntr}
\usepackage{listings}
\usepackage[shortlabels]{enumitem}

\counterwithin{equation}{section}

\newcommand*{\affaddr}[1]{#1} % No op here. Customize it for different styles.
\newcommand*{\affmark}[1][*]{\textsuperscript{#1}}

\def\Cincy{\small{Department of Physics, University of Cincinnati, Cincinnati, Ohio 45221, USA}}
\def\Yale{\small{Department of Physics, Sloane Laboratory, Yale University, New Haven, Connecticut 06520, USA}}

\begin{document}

% Don't want date printed

% Make title large and bold
\title{\Large\bfseries RG flow and symmetry breaking in a weakly coupled model}

\author{
Thomas Appelquist\affmark[$\dag$], Lauren Street\affmark[$\dag\dag$], and L.C.R. Wijewardhana\affmark[$\dag\dag$] \\
{\it\affaddr{\affmark[$\dag$]\Yale}} \\
{\it\affaddr{\affmark[$\dag\dag$]\Cincy}}
}

\begin{abstract}
We explore a weakly coupled $SU(N_{c})$ gauge theory, examining its
fixed-point structure and the transition from infrared conformality to
spontaneous symmetry breaking. Following a previous study, we couple the
gauge field to $N_s$ scalars and $N_f$ fermions, take the large-$N_c$
limit with $N_s/N_c$ and $N_f/N_c$ fixed, and adjust these ratios to
produce weakly coupled infrared fixed points while maintaining
asymptotic freedom. We map out renormalization-group trajectories, and
describe the transition to symmetry breaking as the ratio $N_s/N_c$ is
increased. We examine the breaking by employing the one-loop effective potential. At tree level, a scalar-field vacuum expectation value (VEV) of a certain form is allowed, leading to masses for a subset of the scalar fields. Among the remaining massless scalars, one corresponds to a massless dilaton and the others are Nambu-Goldstone bosons absorbed by the gauge fields. The one-loop potential breaks the scale symmetry explicitly, determining the VEV and generating a dilaton mass suppressed relative to the other masses by a factor of the weak scalar coupling.
\end{abstract}

\maketitle

\section{Introduction}
Gauge theories exhibit different types of infrared (IR) behavior depending on the matter-field content. A much studied example is an $SU(N_{c})$ gauge theory with a set of $N_f$ massless Dirac fermions in the fundamental representation. Confinement and spontaneous chiral symmetry breaking set in at low values of $N_f$, while the theory exhibits conformal symmetry when $N_f$ is near $11 N_c/2$ where asymptotic freedom is lost. The transition between the two takes place at some critical value $N_{fc}$, above which the theory is said to be in the conformal window. The precise value of $N_{fc}$ and the behavior of the theory as $N_f$ approaches $N_{fc}$ from either side are strong-coupling questions.

It can be illuminating to study these issues in a context where a weak-coupling expansion is possible. An interesting such model with an $SU(N_{c})$ gauge symmetry was examined in Ref. \cite{Benini:2019dfy}, building on an earlier treatment \cite{Hansen:2017pwe}, and making use of a $1/N_c$ expansion. Here we provide a complementary study,
expanding on this work. We discuss renormalization-group (RG) evolution,
describing infrared behavior as a function of the parameters of the
theory while maintaining asymptotic freedom. We examine the parametric
boundary between IR conformal behavior and symmetry breaking, and
describe the emergent mass spectrum in full detail.

It has been speculated that the transition from IR conformal behavior to a broken-symmetry phase is governed by the merger of IR and ultraviolet (UV) fixed points \cite{Kaplan:2010zz}. There is no evidence for this in the case of strongly coupled theories such as the $SU(N_{c})$ gauge theory with $N_f$ Dirac fermions, but the analysis of Ref. \cite{Benini:2019dfy} suggested that this does happen in the theory studied there.

The model to be studied is an $SU(N_{c})$ gauge theory with $N_f$ massless Dirac fermions and $N_s$ massless complex scalars, both in the fundamental representation of the gauge group. The masslessness of the fermions is protected by chiral symmetry while the scalar mass must be tuned to zero. We work in the large-$N_c$ limit with the ratios $N_{f}/N_c$ and $N_{s}/N_c$ fixed.

In Sec. \ref{sec:theory}, we describe the model, involving a gauge coupling and two scalar couplings, and record the three coupled renormalization-group (RG) equations in the large-$N_c$ limit.

In Sec. \ref{sec:RGflow}, we study the RG evolution governed by these three equations. We determine qualitative features analytically, and compute certain RG trajectories numerically.  For the ratio $x_s \equiv N_{s}/N_c$ below a certain critical value $\bar{x}_s$, we confirm the existence of generic RG trajectories for which the couplings run to the origin in the UV, and to weak nonzero fixed points in the IR. When  $x_s$ is above $\bar{x}_s$, one of the scalar couplings runs instead to decreasing negative values, triggering spontaneous symmetry breaking and the appearance of Nambu-Goldstone bosons (NGBs). Asymptotic freedom remains for a range of $x_s$ values. We comment on the merging of fixed points as the theory transitions into the broken phase.

In Sec. \ref{sec:symm}, we discuss the pattern of symmetry breaking for $x_s > \bar{x}_s$ by analyzing the one-loop effective potential. We first observe that at an appropriately defined tree level, a scalar-field vacuum expectation value(VEV) of a certain form is allowed. The full set of $2N_{s}N_{c}$ real scalars then breaks into a set of massless NGBs associated with the gauge symmetry breaking, all of which are absorbed by the Higgs mechanism, a single dilaton associated with the spontaneous breaking of the scale symmetry of the tree potential, and a set of massive scalars. We next examine the full effective potential for the dilaton degree of freedom, in which the scale symmetry is broken explicitly.  It dictates the presence of a nonzero VEV and gives rise to a mass for the dilaton. The squared mass is suppressed relative to that of the other massive scalars and the massive gauge bosons by a power of the weak coupling.

In Sec.\ref{sec:concl}, we summarize our results and discuss open questions.

\section{The Model}\label{sec:theory}
The theory is described by the Lagrangian (employing the notation of Ref. \cite{Benini:2019dfy}),
\begin{align}\label{eq:lagrange}
\mathcal{L} &= -\frac{1}{4} F_{\mu\nu} F^{\mu\nu} + \text{Tr} \bar{\psi} i \cancel{D} \psi + \text{Tr} D_\mu \phi^\dagger D^\mu \phi 
\nonumber \\
&\quad- \tilde{h} \text{Tr} \phi^\dagger \phi \phi^\dagger \phi - \tilde{f} \text{Tr} \phi^\dagger \phi \text{Tr} \phi^\dagger \phi,
\end{align}
where $F_{\mu\nu} = F^a_{\mu\nu}t^a = \partial_\mu A_\nu^a t^a - \partial_\nu A_\mu^a t^a$, $D_\mu = \partial_\mu - i g A_\mu^a t^a$, and $\text{Tr}(t^a t^b) = (1/2)\, \delta^{ab}$ for $a,b = 1,...,N_c^2 - 1$. We take the limit $N_c, N_f, N_s \gg 1$, holding fixed the couplings,
\begin{align}\label{eq:scalings}
\lambda = \frac{N_c g^2}{16 \pi^2}, \qquad h =\frac{N_c \tilde{h}}{16 \pi^2}, \qquad f = \frac{N_c N_s \tilde{f}}{16 \pi^2}.
\end{align}
In this limit, the three RG equations are, to leading order, governed by the beta functions  \cite{Benini:2019dfy,Hansen:2017pwe,Sannino:2016sfx,Machacek:1984zw},
\begin{align}
\beta_\lambda &= -\frac{1}{3} \left(22 - x_s - 4 x_f\right) \lambda^2 
\nonumber \\
&\quad+ \frac{2}{3} \left(4 x_s + 13 x_f - 34\right) \lambda^3 \label{eq:betal},
\\
\beta_h &= 4(1+x_s) h^2 - 6 \lambda h + \frac{3}{4} \lambda^2 \label{eq:betah},
\\
\beta_f &= 4 f^2 + 8\left(1 + x_s\right) f h + 12 x_s h^2
\nonumber \\
&\quad - 6 \lambda f + \frac{3 x_s}{4} \lambda^2 \label{eq:betaf},
\end{align}
where
\begin{align}
x_s = N_s/N_c, \qquad x_f = N_f/N_c.
\end{align}

The beta function $\beta_{\lambda}$ (Eq. (\ref{eq:betal})), taken to two loops, depends on only $\lambda$ itself, and is asymptotically free for $x_s + 4x_f < 22$. It exhibits a weak, Caswell-Bank-Zaks IR fixed point when $(22 - x_s - 4x_f)  \ll 2 ( 4x_s + 13 x_f - 34)$. The fixed-point value is
\begin{align}
\lambda^* &= \frac{\epsilon}{1 + x_s/50 - 13 \epsilon/2},
\end{align}
where 
\begin{align}
\epsilon &= \frac{1}{75}\left(22 - x_s - 4 x_f\right) \ll 1.
\end{align}

\section{RG Trajectories}\label{sec:RGflow}
The structure of the beta functions (Eqs. (\ref{eq:betal}) - (\ref{eq:betaf})) allows us to analyze the RG running sequentially, first describing the flow of $\lambda(\mu)$ as function of the RG scale $\mu$, then $h(\mu)$, and then $f(\mu)$. We supplement our analytic study with numerical solutions to the combined RG equations.

\subsection{\boldmath{$\lambda(\mu)$} flow}
\begin{figure}
\centering
\includegraphics[width=0.45\textwidth]{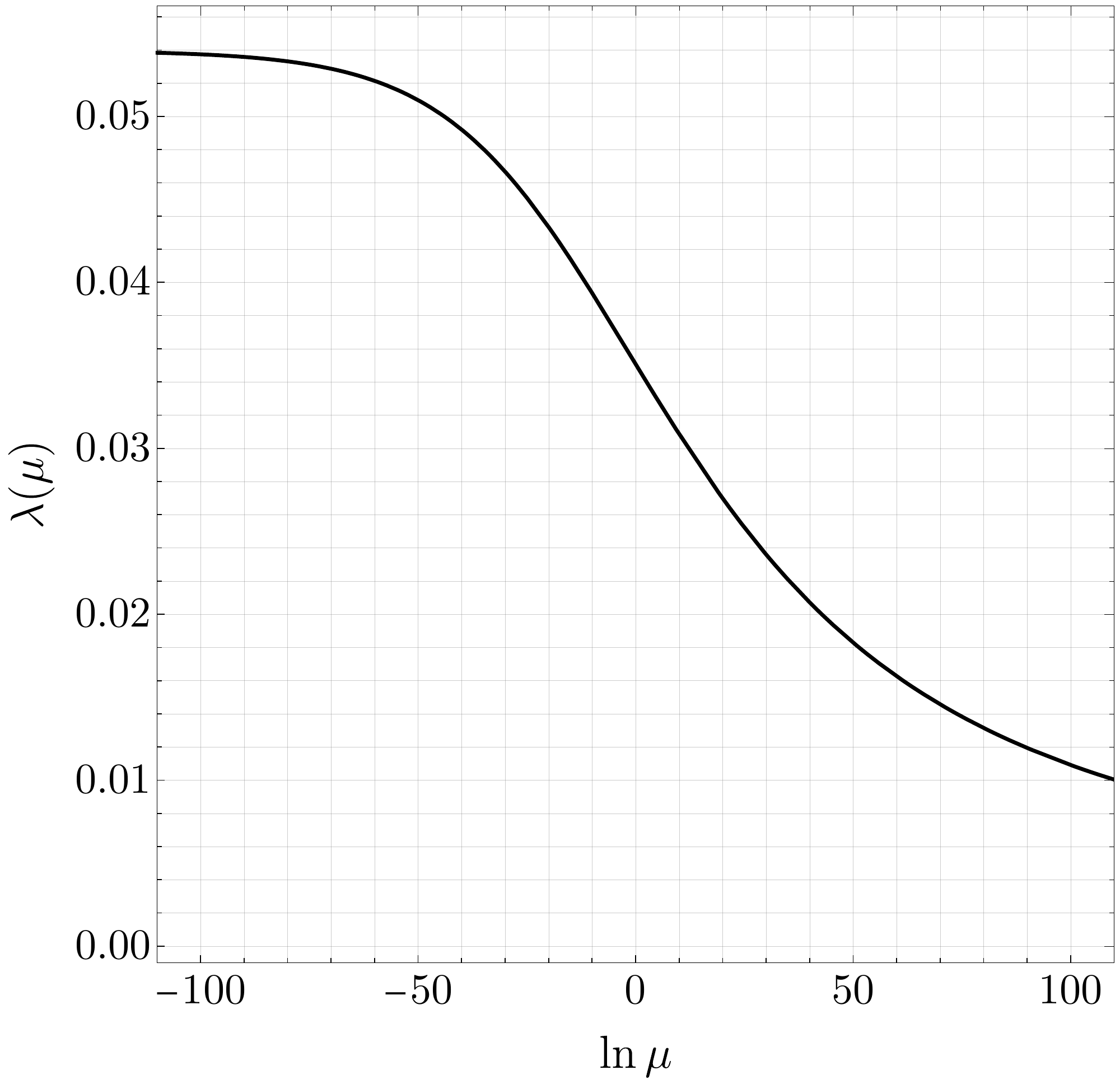}
\caption{Flow of $\lambda(\mu)$ with $\ln\mu$ for $\epsilon = 0.04$
and $x_s = 0.06$. The flow corresponds to an assigned value of $\lambda(\mu)/\lambda^*=0.65$ at the arbitrary scale $\mu = 1$.}
\label{fig:l_vsMu_IR_UV}
\end{figure}

We take the gauge coupling $\lambda(\mu)$ to run in the interval between $0$ and $\lambda^*$ as $\mu$ ranges from $\infty$ to $0$. With $\lambda^* \ll 1$, higher order corrections to $\beta_{\lambda}$ may reliably be neglected throughout this interval. A closed form for the behavior of $\lambda(\mu)$ can be written in terms of the Lambert W function \cite{Gardi:1998qr,Corless_etal}.  In Fig. \ref{fig:l_vsMu_IR_UV}, we display $\lambda(\mu)$ versus $\ln\mu$ generated
numerically, with the parameter choice $x_s = 0.06$ (near the requisite
value for symmetry breaking as we shall discuss). We also take $\epsilon
= 0.04$, a value small enough to ensure the reliability of the IR fixed
point $\lambda^* = 0.054$. We normalize the curve choosing
$\lambda(\mu)/\lambda^* = 0.65$ at the arbitrary scale $\mu = 1$.

\subsection{\boldmath{$h(\mu)$} flow}
The beta function $\beta_h$, taken to one loop, depends on only $h(\mu)$ and $\lambda(\mu)$ (whose behavior is already determined).  As a function of $h(\mu)$, $\beta_h$ has zeros at
\begin{align}\label{eq:zerosh}
h_{\pm}(\mu) &= \frac{\lambda(\mu)}{1+x_s} \left(\frac{3}{4} \pm B\right),
\end{align}
where
\begin{align}
B &= \frac{\sqrt{6 - 3 x_s}}{4}.
\end{align}
which is real and positive for $x_s < 2$.  In this range, as $\mu \rightarrow 0$, the zero-points $h_{\pm}(\mu)$, increase to
\begin{align}\label{eq:hpm}
h_{\pm}^* \equiv \frac{\lambda^*}{1+x_s}\left( \frac{3}{4} \pm B\right),
\end{align} 
The upper zero-point $h_{+}(\mu)$ of Eq. (\ref{eq:zerosh}) is IR attractive for $h(\mu)$ flow and therefore  the pair $(\lambda^*,h_+^*) $ is a stable IR fixed point in the $(\lambda(\mu), h(\mu))$ plane. The lower zero-point $h_{-}(\mu)$ is IR repulsive and therefore $h_-^*$ plays no direct role in our description of the RG trajectories.

To further analyze the RG  flow of $h(\mu)$ with $x_s < 2$, we choose normalizing values of $h(\mu)$ between $h_{-}(\mu)$  and $h_{+}(\mu)$, at the arbitrary reference scale $\mu= 1$. In this range, $\beta_h$ is negative. With the normalizing value not too close to the above boundaries, $h(\mu) \rightarrow h_{+}^{*}$ as $\mu \rightarrow 0$, and $h(\mu) \rightarrow 0$ as $\mu \rightarrow \infty$. If, on the other hand, normalizing values of $h(\mu)$ were chosen too close to the upper boundary, asymptotic freedom would be lost since the decrease of $h_{+}(\mu)$ with increasing $\mu$ would overtake the decrease of $h(\mu)$, turning $\beta_h$ positive. Similarly, normalizing values of $h(\mu)$ too close to $h_{-}(\mu)$ would disrupt the IR behavior, with $h_{-}(\mu)$ overtaking $h(\mu)$ as $\mu$ decreases. This would drive $h(\mu)$ away from the IR fixed point $h_+^*$.
\begin{widetext}
\begin{figure*}[t]
\centering
\begin{subfigure}{0.45\textwidth}
\includegraphics[width=\textwidth]{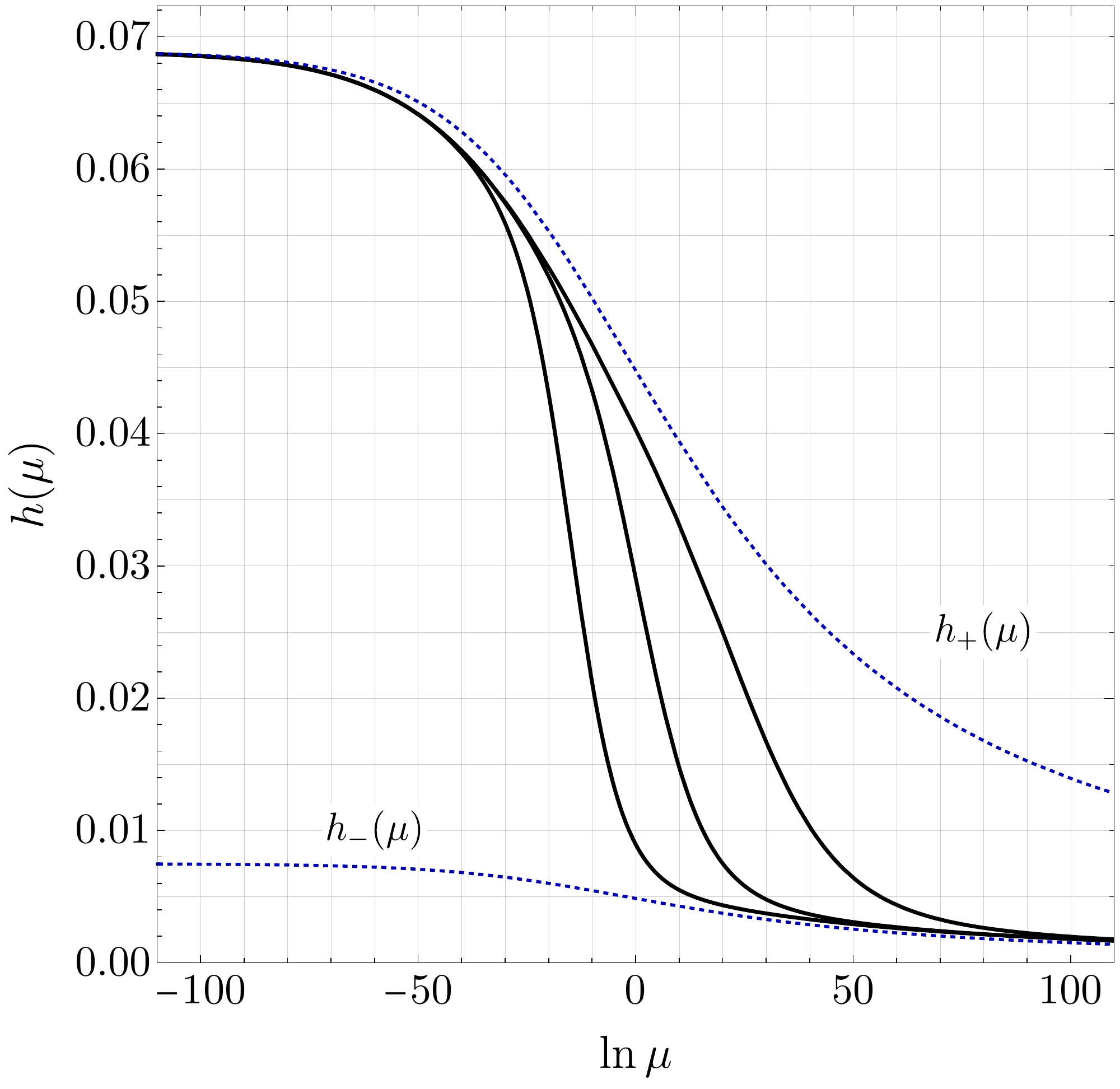}
\caption{Flow of $h$ with $\ln\mu$}
\label{fig:h_vsMu_IR_UV}
\end{subfigure}
\begin{subfigure}{0.45\textwidth}
\includegraphics[width=\textwidth]{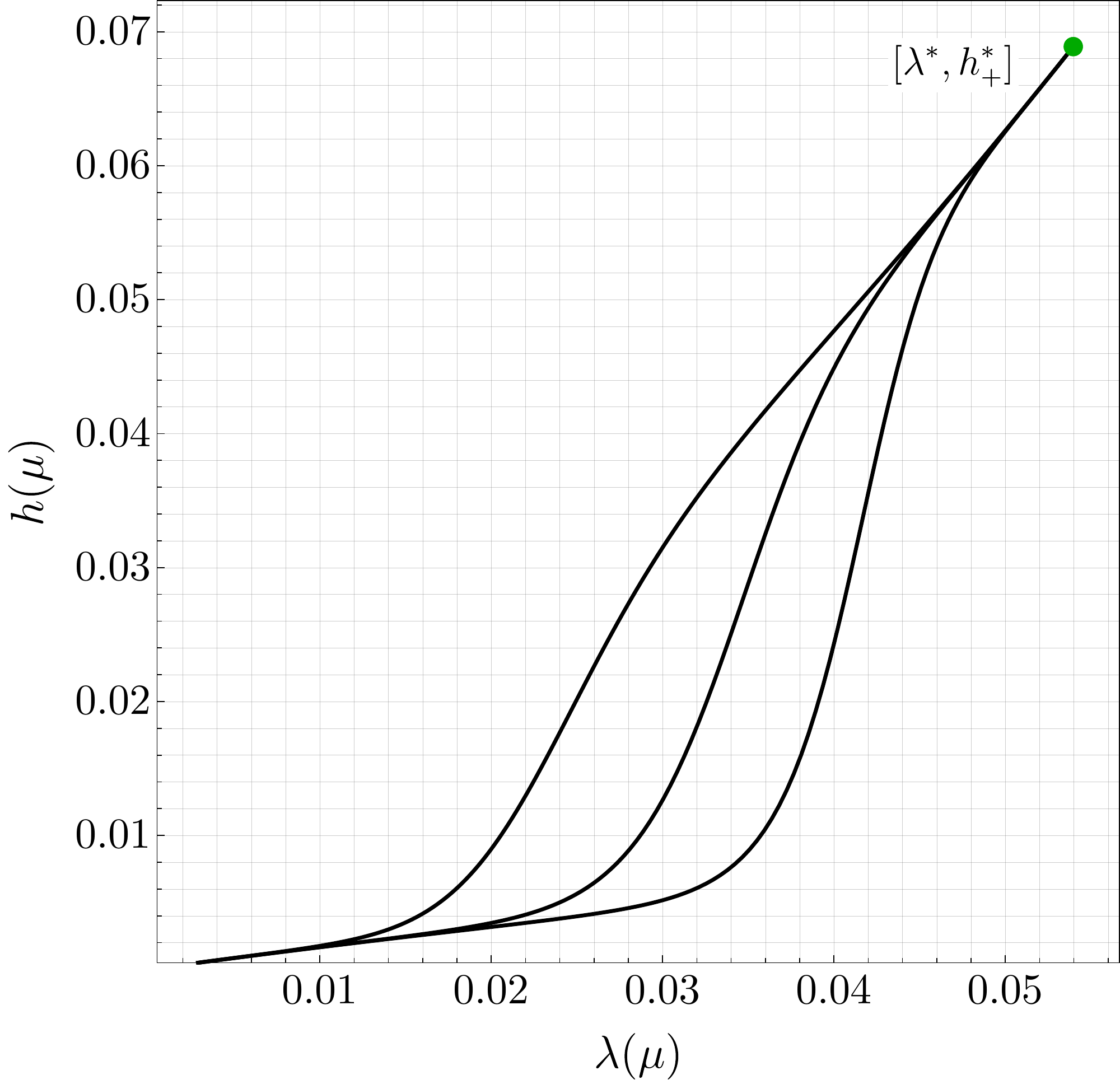}
\caption{Flow in the $\left(\lambda(\mu),h(\mu)\right)$ plane}
\label{fig:lh_IR_UV}
\end{subfigure}
\caption{Flow for $x_s = 0.06$ and $\epsilon = 0.04$.  The RG trajectories are given by the solid black lines and correspond to an assigned value of $\lambda(\mu)/\lambda^* = 0.65$ and $h(\mu)/h_+(\mu) = (0.2, 0.65, 0.9)$ at the arbitrary scale $\mu = 1$.  The upper dotted blue line in Fig. \ref{fig:h_vsMu_IR_UV} corresponds to $h_+(\mu)$, and the lower dotted blue line corresponds to $h_-(\mu)$ where $h_{\pm}(\mu)$ is given by Eq. (\ref{eq:zerosh}).}
\label{fig:hplots}
\end{figure*} 
\end{widetext}

In Table  \ref{tab:hcross} we show, for the parameter choices $\epsilon = 0.04$ and $x_s = 0.06$, allowed upper values of $h(\mu)/h_{+}(\mu)$ at $\mu = 1$ to ensure asymptotic freedom, and
allowed lower values of $h(\mu)/h_{-}(\mu)$ at $\mu = 1$ to ensure the IR dominance of the fixed point $h_{+}^*$. In Fig. \ref{fig:h_vsMu_IR_UV}, we plot three such trajectories generated numerically, again for parameter choices $\epsilon = 0.04$ and $x_s = 0.06$. Each takes $\lambda(\mu)/\lambda^* = 0.65$ at $\mu= 1$ as in Fig.  \ref{fig:l_vsMu_IR_UV}, and one of three different choices for $h(\mu)/h_{+}^{*}$ at the same scale. Fig. \ref{fig:lh_IR_UV} shows the corrsponding two-dimensional flow.

\begin{widetext}
\begin{table*}[t]
\centering
\begin{tabular}{|c|c|c|c|}
\hline
$\epsilon$ & $\lambda(\mu=1)/\lambda^*$ & IR: $h(\mu=1)/h_{-}(\mu=1)$ & UV:  $h(\mu=1)/h_{+}(\mu=1)$ 
\\
\hline \hline
0.04 & 0.65 & $\geq 1.0678$ & $\leq 0.9148$
\\ \hline
\end{tabular}
\caption{Limits on the normalized values for $h(\mu)$ at the scale $\mu = 1$.  The lower limit (third column) is for stability in the IR and is given in units of $h_{-}(\mu)$.  The upper limit (fourth column) is for asymptotic freedom and is given in units of $h_{+}(\mu)$.}
\label{tab:hcross}
\end{table*}
\end{widetext}

\subsection{\boldmath{$f(\mu)$} flow}\label{sec:fflow}
It is here that the transition
between IR conformality and symmetry breaking appears, as a function of $x_s$. The beta function $\beta_f$ (Eq. (\ref{eq:betaf})) has two zeros given by
\begin{widetext}
\begin{align}\label{eq:zerosf}
f_{\pm}(\mu) &=\frac{1}{4} \Bigg(3 \lambda(\mu) - 4 h(\mu)(1+x_s)
\nonumber \\
&\pm \sqrt{16 h(\mu)^2(1+x_s(x_s-1))- 3 \lambda(\mu)^2(x_s - 3) - 24 h(\mu) \lambda(\mu) (1+x_s)}\Bigg)
\end{align}
\end{widetext}
The location of these zero-points, in particular whether they are real, depends on the behavior of $\lambda(\mu)$ and $h(\mu)$.

In the IR, where $(\lambda(\mu), h(\mu)) \rightarrow (\lambda^*, h_{+}^*)$, the zero points are at
\begin{align}
f^*_{\pm} = \lambda^*(- B \pm A_{+}),
\end{align}
where
\begin{align}
A_{\pm} = \frac{3 \sqrt{2 - \left(13 \pm 6 \sqrt{6 - 3x_s}\right) x_s + x_s^2 - 2x_s^3}}{4 \sqrt{3}(1+x_s)}.
\end{align}
Both zero points are real (and negative) for $x_s \leq \bar{x}_s \approx 0.07$, which is the value of $x_s$ above which $ A_{+}$ acquires an imaginary part.  We first discuss RG flow for this case, and then for $x_s \geq \bar{x}_s$.

\subsubsection{RG flow for $x_s \leq \bar{x}_s$}
The zero point $f_{+}^*$ is IR attractive, meaning that for $x_s \leq \bar{x}_s $, $(\lambda^*,h_+^*,f_+^*)$ is an IR stable fixed point in the full three-dimensional coupling space. 

As $\mu$ increases, $\lambda(\mu)$ and $h(\mu)$ decrease (Fig. \ref{fig:lh_IR_UV}). Furthermore $h(\mu)$ decreases relative to $\lambda(\mu)$ so that at some scale the discriminant of $\beta_f$ turns negative and $\beta_f$ becomes completely positive with no real zeros. When $h(\mu)$ drops to $(3/4)[\lambda(\mu)/(1+x_s)]$, the (positive) minimum of $\beta_f$  is at $f(\mu) = 0$. As $\mu$ increases further, $h(\mu)$ drops below $h(\mu) = (3/4)[\lambda(\mu)/(1+x_s)]$, and the minimal point of $\beta_f$ moves to positive values of $f(\mu)$. As $h(\mu)$ continues to decrease relative to $\lambda(\mu)$, the discriminant again turns positive. There, two real zeros of $\beta_f$ reappear at positive values of $f(\mu)$.  The forward motion of the zero-points turns around since $\lambda(\mu)$ and $h(\mu)$ are both decreasing with increasing $\mu$.  Eventually, $h(\mu) = [\lambda(\mu)/(1+x_s)]( 3/4 - B)$ and the two zeros of $\beta_f$ are at
\begin{align}
\lambda(\mu) ( B \pm A_{-}).
\end{align}
Both zero-points are positive and both decrease to zero with increasing $\mu$. The zero at $\lambda(\mu) (B - A_{-})$ is UV attractive for $f(\mu)$ flow.

\begin{widetext}
\begin{figure*}
\centering
\begin{subfigure}{0.45\textwidth}
\includegraphics[width=\textwidth]{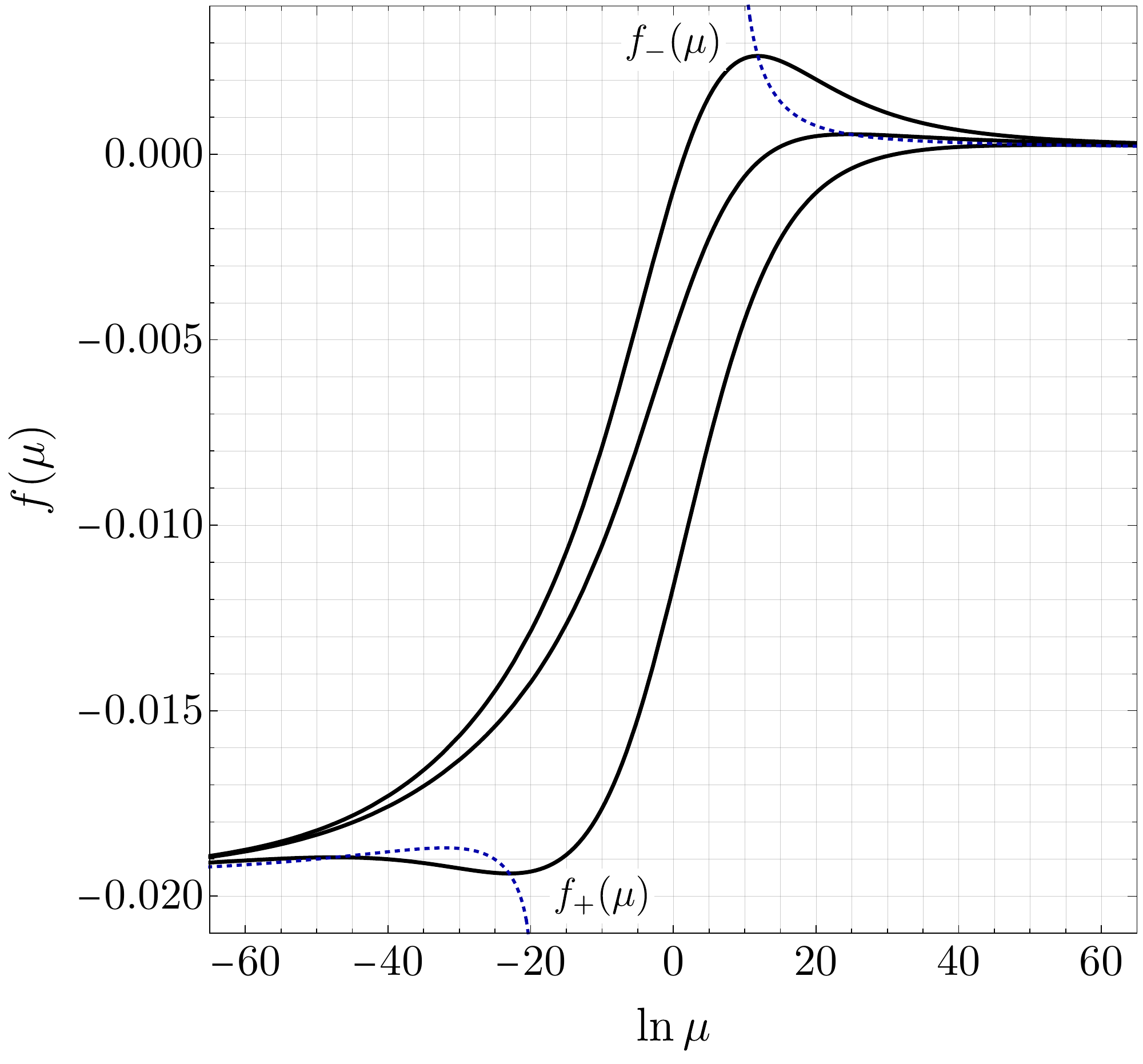}
\caption{Flow of $f$ with $\ln\mu$}
\label{fig:f_vsMu_IR_UV}
\end{subfigure}
\begin{subfigure}{0.45\textwidth}
\includegraphics[width=\textwidth]{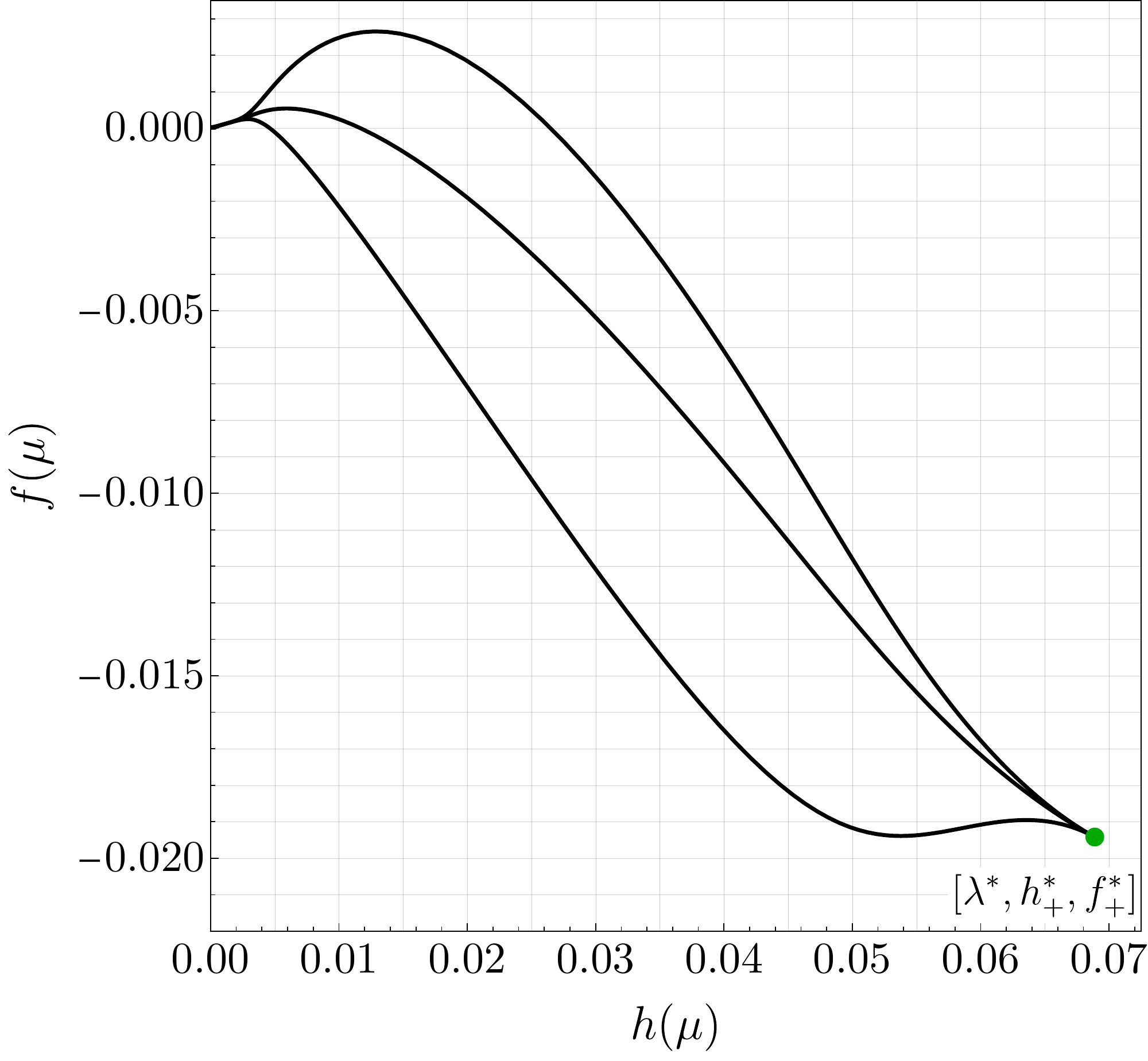}
\caption{Flow in the $\left(h(\mu),f(\mu)\right)$ plane}
\label{fig:hf_IR_UV}
\end{subfigure}
\caption{Flow for $x_s = 0.06$ and $\epsilon = 0.04$.  The RG trajectories are given by the solid black lines and correspond to assigned values $\lambda(\mu)/\lambda^* = 0.65$, $h(\mu)/h_+(\mu) = 0.65$, and $f(\mu)/f_+^* = (0.05,0.25,0.6)$ at the arbitrary scale $\mu = 1$.  The left dotted blue line in Fig. \ref{fig:f_vsMu_IR_UV} corresponds to $f_+(\mu)$, and the right dotted blue line corresponds to $f_-(\mu)$ where $f_{\pm}(\mu)$ is given by Eq. (\ref{eq:zerosf}).}
\label{fig:fplots}
\end{figure*} 
\end{widetext}

This evolving landscape of $\beta_f$ determines the flow of $f(\mu)$, with the detailed trajectory depending on assigned normalizing values for $\lambda(\mu)$, $h(\mu)$, and $f(\mu)$. We choose normalizing values for $f(\mu)$ in a range between $f_{+}^*$ and zero such that $f(\mu)$ flows down to $f_{+}^*$ in the IR, and up through positive values in the UV.  It then decreases to zero being attracted by the decreasing zero point of $\beta_f$ at $\lambda(\mu) (B - A_-)$. This is illustrated in Fig. \ref{fig:fplots} using parameter values $\epsilon = 0.04$ and $x_s = 0.06$. We have also taken $\lambda(\mu)/\lambda^* = 0.65$ at $\mu = 1$ , $h(\mu)/h_+^* = 0.65$ at
$\mu = 1$, and employed three different normalizing values for $f(\mu)$. Fig. \ref{fig:f_vsMu_IR_UV} shows the flow of $f(\mu)$ vs $\ln \mu$ while Fig. \ref{fig:hf_IR_UV} shows the flow in the $(h,f)$-plane.

Each of the RG trajectories of Fig. \ref{fig:fplots} exhibits IR conformality
and asymptotic freedom. The behavior is generic, true for a
range of normalizing values for $f(\mu)$. The allowed range of
normalizing values can be described in a manner parallel to $h(\mu)$ by
specifying $f(\mu)$ relative to $f_{-}(\mu)$ and $f_{+}(\mu)$. We do
this at $\mu$ values such that these quantities are real.  This range, illustrated by the blue dashed lines in Fig. \ref{fig:f_vsMu_IR_UV}, does not include $\mu =
1$, which was used to normalize $\lambda(\mu)$ and $h(\mu)$. In Table \ref{tab:fcross},
we show the upper bound on $f(\mu)/f_{-}(\mu)$ and the lower bound on
$f(\mu)/f_{+}(\mu)$, each at a convenient scale where these quantities
are real.
\begin{table}[ht]
\centering
\begin{tabular}{|c|c|c|c|c|}
\hline
$\epsilon$ & $\lambda(\mu)/\lambda^*$ & $h(\mu)/h_{+}(\mu)$ & IR: $f(\mu)/f_{-}(\mu)$ & UV:  $f(\mu)/f_{+}(\mu)$
\\
\hline \hline
0.04 & 0.65 & 0.65 & $\geq 1.0561$  & $\leq0.8514$
\\ \hline
\end{tabular}
\caption{Lower limit on $f(\mu)$ in units of $f_{-}(\mu)$ for stability in the IR (third column) and upper limit on $f(\mu)$ in units of $f_{+}(\mu)$ for asymptotic freedom (fourth column) assuming $\epsilon = 0.04$, $\lambda(\mu)/\lambda^* = 0.65$, and $h(\mu)/h_+(\mu) = 0.65$.  The assigned values for $\lambda$ and $h$ are at $\ln\mu=0$ while that for the $f$ flow is taken at a different value of $\mu$ in order to ensure real values for $f_\pm(\mu)$.  For the IR stability limit (third column), $f(\mu)$ is assigned at the scale $\ln \mu = -45$, while for the UV stability limit (fourth) column, $f(\mu)$ is assigned at the scale $\ln \mu = 45$.}
\label{tab:fcross}
\end{table}

\subsubsection{RG flow for $x_s > \bar{x}_s$}
When $x_s$ is increased above $\bar{x}_s$, the system undergoes a phase transition from IR conformality to a symmetry-breaking state. Asymptotic freedom is maintained provided $x_s \leq 0.84$ where $A_{-}$ is real, but the range of normalizing values for $f(\mu)$ decreases near the upper bound.  We stay well below the bound, maintaining $x_s \leq 0.1$.  The transition at $\bar{x}_s$ sets in because in the IR, $\beta_f$ no longer has real zeros as a function of $f(\mu)$. It is strictly positive. Thus, while $\lambda(\mu) \rightarrow \lambda^*$ and $h(\mu) \rightarrow h_{+}^*$ in the IR, $f(\mu)$ evolves to ever decreasing negative values. 

As $\mu$ increases, the minimal point of $\beta_f$ increases to positive values of $f(\mu)$ as $h(\mu)$ drops below $(3/4)[\lambda(\mu)/(1+x_s)]$. The further evolution of $\beta_f$ as $\mu$ increases is similar to its behavior for $x_s < \bar{x}_s$, with two real zeros appearing, both of which eventually decrease to zero since $h(\mu)$ and $\lambda(\mu)$ are decreasing with increasing $\mu$.

\begin{widetext}
\begin{figure*}
\centering
\begin{subfigure}{0.45\textwidth}
\includegraphics[width=\textwidth]{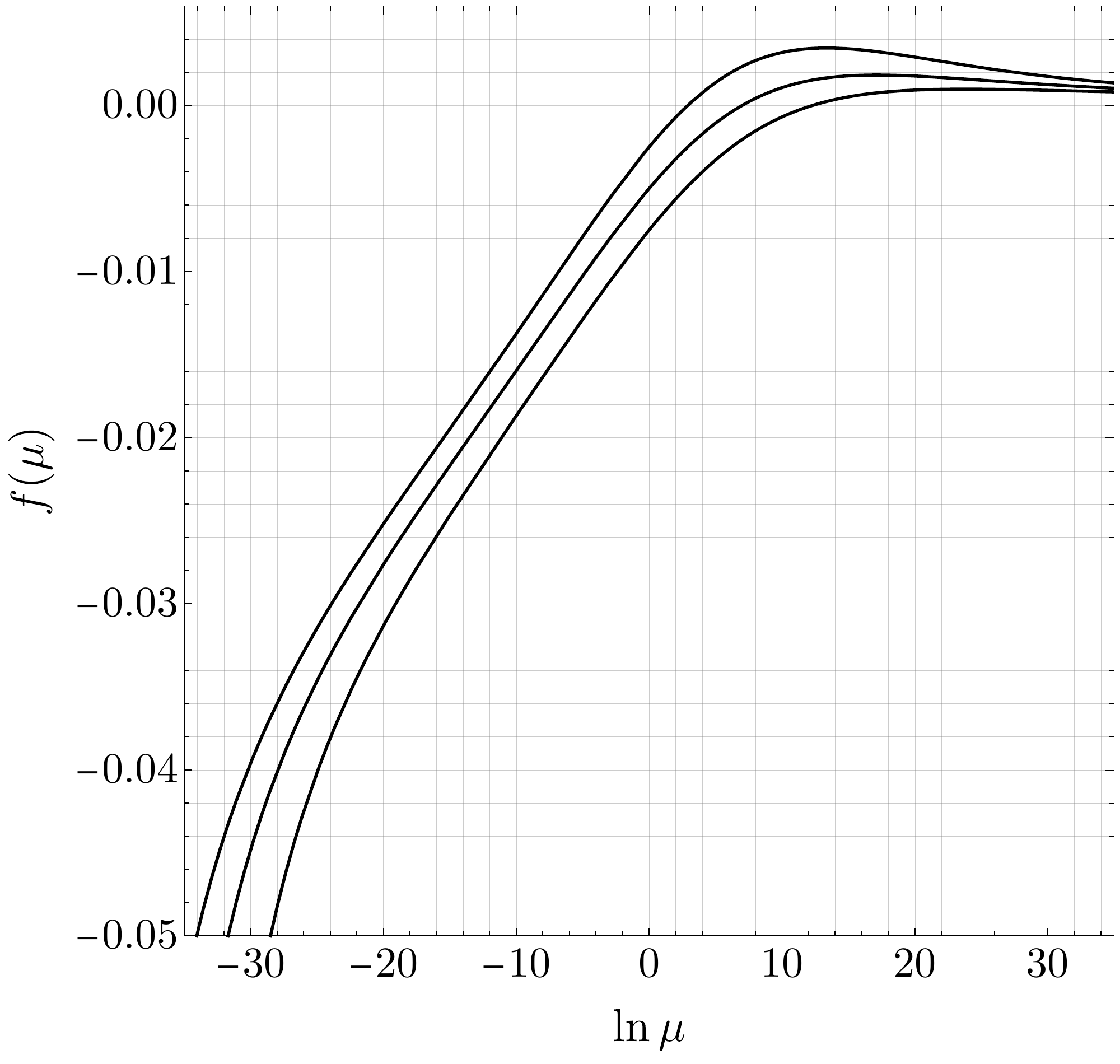}
\caption{Flow of $f(\mu)$ for $x_s = 0.1$}
\label{fig:f_xs_large}
\end{subfigure}
\begin{subfigure}{0.45\textwidth}
\includegraphics[width=\textwidth]{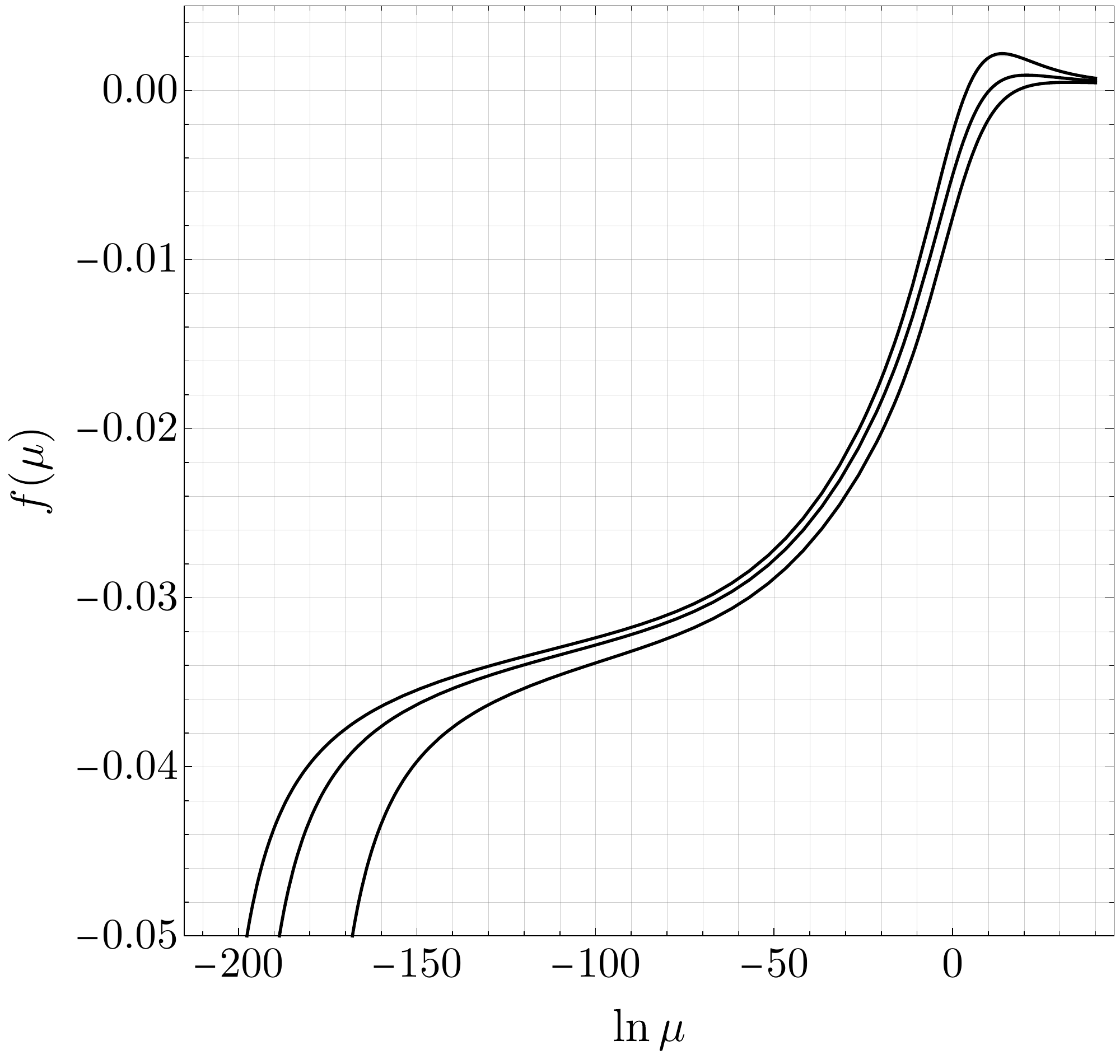}
\caption{Flow of $f(\mu)$ for $x_s = 1.012 \bar{x}_s$}
\label{fig:f_xs_med}
\end{subfigure}
\caption{Flow for $x_s > \bar{x}_s$, $\epsilon = 0.04$, and assigned values $\lambda(\mu)/\lambda^*=0.65$, $h(\mu)/h_+(\mu)=0.65$, and $f(\mu) = (-0.0025,-0.005,-0.0075)$ at the arbitrary scale $\mu = 1$. For $x_s =0.1$ (Fig. \ref{fig:f_xs_large}), $f$ flows to decreasing negative values smoothly, while for $x_s - \bar{x}_s \ll 1$ (Fig. \ref{fig:f_xs_med}), $f$ plateaus before decreasing further.}
\label{fig:fplots_xslarge}
\end{figure*}
\end{widetext}

The behavior of $f(\mu)$ is illustrated in Fig. \ref{fig:fplots_xslarge} for the case $\epsilon = 0.04$. In Fig. \ref{fig:f_xs_large}, we take $x_s = 0.1$, a value above $\bar{x}_s$ (outside the conformal window). The nonmonotonic flow from zero in the UV to decreasing negative values in the IR is evident. In Fig. \ref{fig:f_xs_med}, we take $x_s = 0.074$ a value just above $\bar{x}_s$ (just outside the conformal window). Now $f(\mu)$ evolves slowly over a large $\mu$ range, leading to a
hierarchy between a scale $\Lambda_{UV}$ where asymptotic freedom sets
in and a scale $\Lambda_{IR}$ where $f(\mu)$ descends to large enough
negative values to trigger spontaneous breaking \footnote{This role of the coefficient of the ``double-trace" operator in the Lagrangian was emphasized in a related model in Ref. \cite{Pomoni:2009joh}.}.

\subsection{Summary of RG flow}
In the $x_s$-parameter range of interest, the theory is asymptotically free. There is a class of RG trajectories for which $\lambda(\mu)$,
$h(\mu)$ and $f(\mu)$ flow in the UV to the origin in the
three-dimensional coupling space. Within this range, however,
an IR transition takes place at $x_s = \bar{x}_s \approx 0.07$.

For $x_s < \bar{x}_s$, the theory is inside the conformal window, with
RG trajectories flowing in the IR to the stable  fixed point
$(\lambda^{*}, h_+^{*}, f_+^{*})$. The evolution of $\lambda(\mu)$  and
$h(\mu)$ along these trajectories is monotonic, as illustrated in Figs.
\ref{fig:l_vsMu_IR_UV} and \ref{fig:hplots}. The trajectories for $f(\mu)$ are non-monotonic, as illustrated in Fig. \ref{fig:fplots}.  We have chosen parameters such that each of these couplings
is weak at all points along the trajectories, justifying the truncation
of  $\beta_{\lambda}$ at two loops, and $\beta_{h}$ and $\beta_{f}$
at one loop.

For $x_s > \bar{x}_s$, the IR trajectories change. It remains the case
that $\lambda(\mu) \rightarrow \lambda^*$ and $h(\mu) \rightarrow h_+^*$,
but $f(\mu)$ evolves to decreasing negative
values, as illustrated in Fig. \ref{fig:fplots_xslarge}. The reason is that in the IR, the
evolving $\beta_f$ no longer has real zeros as a function of $f(\mu)$.
It is strictly positive. This behavior is signaled as $x_s \rightarrow
\bar{x}_s$ from below by the merging of the IR stable fixed
point at $(\lambda^*, h_+^*, f_+^*)$ with an unstable fixed point at
$(\lambda^*, h_+^*, f_-^*)$ where $f_-^* = \lambda^{*}(-B -A_{+})$. The
latter plays no direct role in determining the RG trajectories of interest.

The IR evolution of $f(\mu)$ to decreasing negative values triggers the
spontaneous breaking of certain symmetries and the formation of
Nambu-Goldstone bosons. We turn next to a discussion of symmetry
breaking.

\section{Symmetry-Breaking Pattern}\label{sec:symm}
The Lagrangian (Eq. (\ref{eq:lagrange})) has a local $SU(N_{c})$ symmetry, a global $U(N_{s})$ symmetry associated with the scalars, and a global $SU(N_{f}) \times SU(N_{f} ) \times U(1)$ symmetry associated with the massless fermions. With proper regularization, this symmetry remains at the quantum level. In addition, having set the masses to zero, the model is scale invariant at the classical level. Having specified the scalar couplings at the RG scale $\mu$, we find it helpful to use these couplings to define a ``tree" scalar potential at this scale:
\begin{align}
V_0 = \tilde{h}(\mu) \text{Tr} \phi^\dagger \phi \phi^\dagger \phi + \tilde{f}(\mu) \text{Tr} \phi^\dagger \phi \text{Tr} \phi^\dagger \phi \,,
\end{align}
where $\phi$ is a complex $N_{c} \times N_{s}$ field, and where $\tilde{h}(\mu)$ and $\tilde{f}(\mu)$ are related to $h(\mu)$ and $f(\mu)$ by Eq. (\ref{eq:scalings}).

\begin{widetext}
\begin{figure*}
\centering
\begin{subfigure}{0.45\textwidth}
\includegraphics[width=\textwidth]{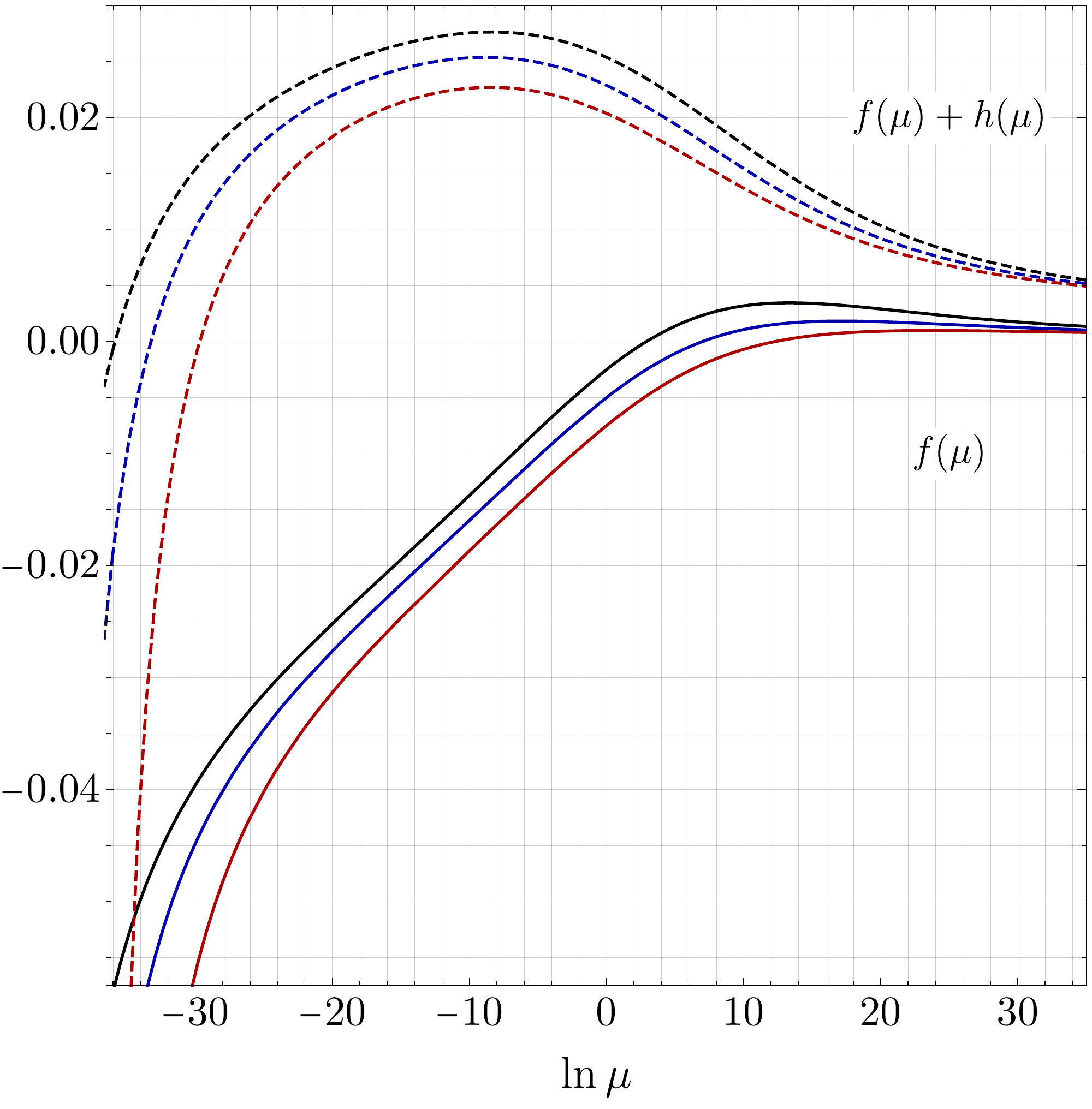}
\caption{Flow of $f(\mu)$ (solid) and $f(\mu)+h(\mu)$ (dashed) for $x_s \approx 1.4 \bar{x}_s$}
\label{fig:fplush_xs_large}
\end{subfigure}
\begin{subfigure}{0.45\textwidth}
\includegraphics[width=\textwidth]{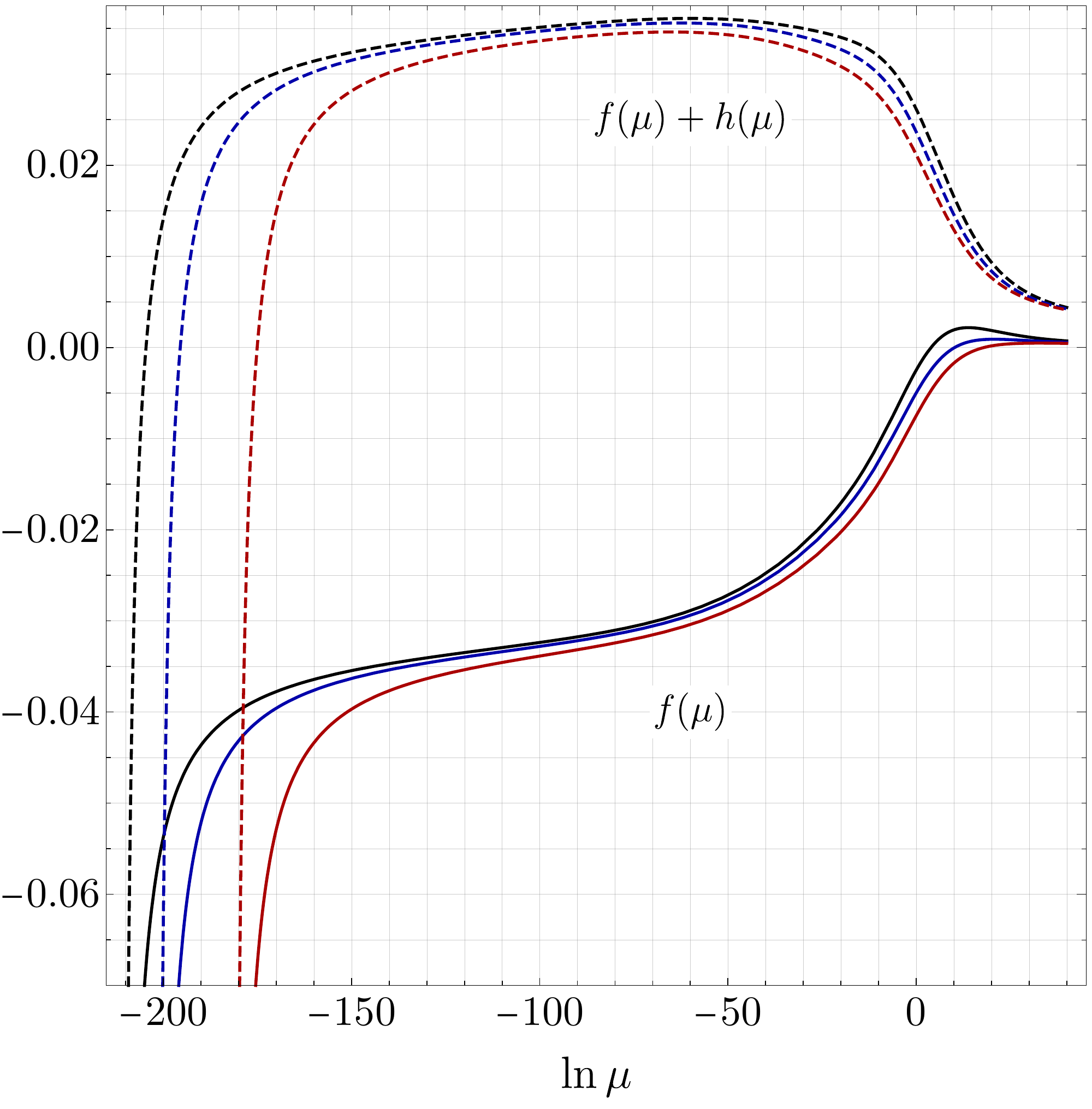}
\caption{Flow of $f(\mu)$ (solid) and $f(\mu)+h(\mu)$ (dashed) for $x_s = 1.012 \bar{x}_s$}
\label{fig:fplush_xs_med}
\end{subfigure}
\caption{Flow of $f(\mu)$ and $f(\mu)+h(\mu)$ for the same values as in Fig. \ref{fig:fplots_xslarge}.  When $f(\mu)+h(\mu)=0$, $f \approx -h_+^*$}
\label{fig:fplushplots_xslarge}
\end{figure*}
\end{widetext}

For $x_s > \bar{x}_s$, the IR evolution of $f( \mu)$ is to decreasing negative values, while $h( \mu) \rightarrow h_ +^* > 0$ and $\lambda( \mu) \rightarrow \lambda^* > 0$. This drives the quantity $h(\mu ) + f(\mu)$, which plays a key role in symmetry
breaking, to small and then negative values \footnote{The flow of $h(\mu)+f(\mu)$ to small and then negative values appears only for $x_s > \bar{x}_s$ because we have restricted attention to RG trajectories that are asymptotically free. For a discussion of symmetry breaking in related models, but without this restriction, see Refs. \cite{Grinstein:2011dq,Antipin:2011aa,Antipin:2012sm}.}.  This flow is shown in Fig. \ref{fig:fplushplots_xslarge} using the same
values for $\epsilon$ and $x_s$ as in Fig. \ref{fig:fplots_xslarge} (see also Ref. \cite{Benini:2019dfy}).  An analysis of symmetry breaking can be carried out using any reference scale for the definition of the couplings, but it is convenient to work at the scale $\Lambda_{IR}$ where $h(\Lambda_{IR}) + f(\Lambda_{IR}) = 0$. At this scale, it can be seen from Fig. \ref{fig:fplushplots_xslarge} that for these trajectories, $h( \Lambda_{IR}) \approx h_ +^*$.

Symmetry breaking is signaled by the appearance of a nonzero vacuum value for the scalar field. It is driven by quantum effects, described here by a one-loop effective potential. This leads to a global minimum with the VEV given by the ``color-flavor-locked" expression $\langle\phi_{i,j}\rangle = v \delta_{ij}$ \cite{Gildener:1975cj},\cite{Hansen:2017pwe}, with ($i,j$) the gauge and flavor indices and $v$ of order $\Lambda_{IR}$. We explore the breaking by first examining the tree potential describing fluctuations about this VEV. Details are provided in the Appendix. With the scale choice $\Lambda_{IR}$, the tree potential is flat and vanishing in the color-flavor-locked direction. Fluctuations about the VEV in this direction then describe a massless particle. This is the dilaton. We then turn to the one-loop effective potential to confirm that $v$ is indeed non-zero, and to compute the mass of the dilaton.

\subsection{Tree potential}
Our interest here is in the range $\bar{x}_s < x_s \leq 0.1$, but for any $N_c$ above $N_s+1$, the symmetry-breaking pattern arising from $ \langle \phi_{i,j}\rangle = v \delta_{ij}$ is \cite{Hansen:2017pwe},
\begin{align} \label{eq:symm_break}
SU(N_c) \times U(N_s) \rightarrow SU(N_c - N_s) \times SU(N_s) \times U(1)\,.
\end{align}
The number of broken generators, and therefore the number of NGBs, is $2 N_c N_s - N_s^2$.  Along with the dilaton, the total number of massless scalar modes is thus expected to be $2 N_c N_s - N_s^2 +1$.  Although our study of RG flow is restricted to leading order in
the large-$N_c$ expansion, here we maintain a precise count of the
various degrees of freedom. We also note that with $N_c$ and $N_s$
restricted to integral values, our discussion of RG flow with
$x_s \equiv N_s/N_c$ very close to the transition value $\bar{x}_s
\approx 0.07$ requires a large value of $N_c$ ($1/N_c \ll \epsilon$).

In the Appendix, we analyze the tree potential with the choice  $f(\mu) + h(\mu) = f(\Lambda_{IR}) + h(\Lambda_{IR}) \equiv 0$, confirming the above count and determining the mass of the remaining $N_s^2 - 1$ scalars. They are degenerate in mass, lying in the adjoint representation of the unbroken $SU(N_s)$ global symmetry. We find,
\begin{align}
m_S^2 = \frac{4 h_+^* (4 \pi v)^2}{N_c},
\end{align}
where we have taken $h \approx h_+^*$.

The $2N_{c}N_{s}  - N_s^2$ NGBs associated with the breaking of $SU(N_{c}) \times U(N_{s})$ are  massless (and remain massless to all orders by the Goldstone theorem). They are absorbed by the gauge bosons that become massive with the breaking of the gauge symmetry $SU(N_c)$ to $SU(N_c - N_s)$.  The resultant squared gauge-boson masses, discussed in the Appendix, are given by,
\begin{align}
m_A^2 = C\lambda^{*}(4 \pi v)^{2}/N_c,
\end{align}
where $\lambda^* \approx h_+^*$ (Eq. (\ref{eq:hpm})), and where $C$ is a constant of order unity. Specifically, $C=1$ for the $N_s^2 -1$ gauge bosons whose generators are nonzero in only the upper $N_s \times N_s$ block, $C=1/2$ for the $2N_{s}(N_c - N_s)$ gauge bosons whose generators are nonzero in only the off-diagonal $N_s \times (N_c-N_s)$ block, and $C = (N_c-N_s) / 2 N_c$ for remaining gauge boson, whose generator is diagonal and traceless, with elements proportional to the identity matrix in each of the two blocks.

\subsection{One-loop potential}\label{sec:1loop}
To demonstrate that the VEV $\langle\phi_{i,j}\rangle = v \delta_{ij}$ is a stable minimum, when $x_s > \bar{x}_s$, we next examine the one-loop effective potential in this direction. With scale symmetry broken explicitly at one loop, this will determine the mass of the dilaton. For simplicity, we replace the external field by its VEV, taking $\phi_{i,j}= v \delta_{ ij}$, where $ i, j \leq N_s$. We work at the RG scale $ \mu = \Lambda_{IR}$ where $h(\Lambda_{IR})+f(\Lambda_{IR}) = 0$.

If this computation is done using a reference scale other than $\Lambda_{IR}$, the one-loop potential  corrects the tree-level contribution $(h(\Lambda) + f(\Lambda))x_{s}(4\pi)^{2}v^4$. The correction is quadratic in the couplings and proportional to $(4\pi)^{2}v^{4} \ln(v^{2}/\Lambda^{2})$. With $h(\Lambda)+f(\Lambda )$ sufficiently small, say of order $h^{2}(\Lambda)$, the full potential can reliably be used to determine the minimum \cite{Coleman:1973jx,Yamagishi:1981qq}. The tree term can then be absorbed into $\Lambda$, replacing it by $\Lambda_{IR}$ , the scale at which $h(\Lambda_{IR}) + f(\Lambda_{IR})$ vanishes. It is essential here that $h(\Lambda)+f(\Lambda)$ be able to \textit{reach} small values relative to $h(\Lambda)$ and other couplings at some scale. Such values are not reached when $x_s < \bar{x}_s$ since $h(\Lambda) + f(\Lambda) \geq h_+^* + f_+^*$ which is of the same magnitude as $h_+^*$ and the other fixed-point couplings. An analysis of the full one-loop potential in this case confirms that it behaves like $(h_+^* + f_+^*) x_{s} (4\pi)^{2}v^4$ as $v\rightarrow 0$. The minimum is at $v=0$ when $x_s < \bar{x}_s$, and the symmetry is unbroken.

To compute the one-loop dilaton potential with $x_s > \bar{x}_s$, one sums over all graphs with external zero-momentum, two-dilaton insertions of $v^{2}$. With $h(\Lambda_{IR})+f(\Lambda_{IR}) = 0$ there is no coupling of this quantity to
the dilaton field itself. In addition, at the quadratic level there is no coupling of the two-dilaton
insert to the scalars identified as Nambu-Goldstone fields when symmetry breaking sets in.
Also at this level, the insert does not couple to gauge bosons transforming under the unbroken
$SU(N_{c} - N_{s})$ subgroup of $SU(N_c)$, nor to the (massless) fermions. The contributing quantum
fields are those describing only massive particles after symmetry breaking set in. Summing over these and working to leading order in $1/N_c$, one obtains,
\begin{align}\label{eq:veff_v}
V(v) = 8 \pi^2 x_s \left[ 8 x_s h_+^{*2} + \frac{3}{4} (1 + x_s)\lambda^{*2}\right]v^4 \ln \frac{v^2}{\Lambda_\text{IR}^2}\,,
\end{align}
agreeing with Eq. (20) of Ref. \cite{Benini:2019dfy}.  We have set $h= h_+^*$ and $\lambda = \lambda^*$ with $h_+^*$ and $\lambda^*$ related by Eq. (\ref{eq:zerosh}).  The potential minimum is at $v^2 = e^{-1/2} \Lambda_{IR}^2$.

From this expression, we can determine the mass of the dilaton, remembering to divide by a factor of $N_s$ to properly normalize the dilaton kinetic term. We have
\begin{align}
m_d^2 = 4 (4 \pi v) ^{2} \left[ 8 x_s h_+^{*2} + \frac{3}{4} (1 + x_s)\lambda^{*2}\right] /N_c \,.
\end{align}
This expression is suppressed relative to the scalar squared mass $m_s^2$ and the gauge-boson squared mass $m_A^2$ by an additional factor of the small coupling constant $h_+^{*}$ ($\approx \lambda^*$). With the numerical values employed in Sec.\ref{sec:RGflow}, it is light relative to the other massive scalars and the massive gauge bosons.  A discussion of this type of suppression was provided by Gildener and Weinberg \cite{Gildener:1976ih}.

The potential $V (v)$ and the dilaton squared mass are proportional to
 $\beta_h +  \beta_f$ in the limit $h+f = h(\Lambda_{IR})+f(\Lambda_{IR}) = 0$. (We have also set
$h(\Lambda_{IR}) = h_+^*$  and $\lambda(\Lambda_{IR}) = \lambda^*$). It is necessary for stability that the sum $\beta_h +\beta_f$ be positive as shown in Eq. (\ref{eq:veff_v}), although at scales $\Lambda$ well above $\Lambda_{IR}$ the beta functions and their sum are negative. Still, the coefficient of the  one-loop term in $V(v)$, as defined in Ref. \cite{Coleman:1973jx}, remains positive when defined at the higher scales where $h+f >0$. It is composed of the sum of the beta functions but with the cross term proportional to $\lambda(h + f)$ omitted \cite{Hill:2014mqa}.(The cross term arises purely from wave-function renormalization in the Landau gauge.)  We note also that since $\beta_ h = 0$ when $ h = h_+^*$ , the potential and the dilaton squared mass are approximately proportional to just $\beta_f$ , which encodes all the breaking of scale symmetry when $h = h_+^*$  and $\lambda = \lambda^*$.

To summarize, symmetry breaking with $N_c \geq N_s +1$ leaves us with 
\begin{enumerate}
\item $N_s^2 -1$ massive scalars with squared masses $m_S^2 = 4 h_+^{*} (4 \pi v)^{2} / N_c$, in the adjoint representation of the unbroken global symmetry group $SU(N_s)$
\item One massive scalar (the dilaton) with a squared mass of order $4h_+^{*2} (4 \pi v)^{2} / N_c$ (suppressed by the factor $h_+^*$)
\item $2N_{c}N_s - N_s^2$ massive vector bosons with squared masses
of order $\lambda^* (4 \pi v)^{2}/N_c$
\item $(N_c - N_s)^2 -1$ massless gauge bosons in the adjoint
representation of the unbroken gauge group $SU(N_c-N_s)$
\item $N_{f} N_c$ massless Dirac fermions.
\end{enumerate}

Each of the nonzero masses is suppressed by a factor of $1/\sqrt{N_c}$. The dilaton squared mass is further suppressed
relative to others by a factor of $h_+^*$. Finally, we note that there is no
further relative suppression of the dilaton mass in the ``walking" limit
$x_s \rightarrow \bar{x}_s$ from above. While $v \approx \Lambda_{IR}$ \textit{is} suppressed in this limit relative, say, to the UV scale where asymptotic freedom sets in, the IR scale $v$ is common to all the masses.

\section{Summary and Conclusion}\label{sec:concl}
We have studied RG flow and the onset of
symmetry breaking in a weakly coupled model. An $SU(N_c)$ gauge field is coupled to a set of $N_f$ massless fermions and $N_s$ massless scalars \cite{Benini:2019dfy,Hansen:2017pwe}, both in the fundamental representation of the gauge group. Weakness is achieved by taking the large-$N_c$ limit with $x_f = N_f/N_c$ and $x_s=N_s/N_c$ fixed and adjusting these ratios. With $\epsilon \equiv (22 - x_s - 4x_f)/75 \ll 1$, and $x_s$ below a critical value $\bar{x}_s \approx 0.07$, this leads to a weak infrared-stable fixed point in the three-dimensional coupling space. We described the RG contours noting the nonmonotonic flow in one of the three directions.

When $x_s$ exceeds $\bar{x}_s$, the infrared fixed point disappears
and a transition to a symmetry-breaking phase takes place while
maintaining asymptotic freedom. The breaking, signaled by a nonzero
vacuum value $v$ of the scalar field, is driven by a one-loop effective
potential. We have described the symmetry-breaking pattern, giving rise
to a set of NGBs with the spontaneous
breaking of the scale symmetry. Masses of the remaining scalars are
computed, with the dilaton mass suppressed
by a factor of the weak coupling. The fermions remain massless
throughout. With an appropriate choice for the RG scale defining the
couplings, each of the masses other than that of the dilaton can be
computed from the tree potential.  An important feature of
the mass spectrum is that each is further suppressed relative to $ 4\pi
v$ by the factor $1/\sqrt{N_c}$.

We have described several aspects of the transition at $x_s = \bar{x}_s$. The
disappearance of the stable infrared fixed point as $\bar{x}_s$ is
approached from below is signaled by the merging of this fixed point
with another, unstable fixed point. Within the broken phase, as
$\bar{x}_s$ is approached from above, a walking feature emerges, with
a hierarchy developing between the vacuum value $v$ and the much larger
scale at which asymptotic freedom sets in. It does not lead to an
enhanced hierarchy between the dilaton and the other scalars.

To conclude, we note that while we have analyzed the RG flow to leading (zeroth) order in the $1/N_c$ expansion, the squared masses that appear in the broken phase ($x_s > \bar{x}_s$) are of order $(4 \pi v)^{2}/N_c$. In the far IR, as the RG scale $\mu^2$ becomes of this order, an analysis to order $1/N_c$ must be carried out. The flow of the double-trace coupling $f$, which continued unabated over a large range, is ultimately  interrupted, with the massive modes decoupling at lower scales. The further IR flow depends on the size $N_c - N_s$ of the unbroken gauge group and the number $N_f$ of massless fermions in its fundamental representation. The appearance of the small masses in the broken phase also indicates that the IR fixed points in $\lambda$ and $h$ could be only approximate. Working to next order in $1/N_c$, these couplings could ultimately flow away from the fixed-point values in the far IR. Our study of Coleman-Weinberg symmetry breaking for $ x_s > \bar{x}_s$ can be systematically corrected to include these $1/N_c$ effects. There remains much to study here, but we have been consistent in noting the existence of the phase transition at $x_s = \bar{x}_s$ to zeroth order in $1/N_c$, and carefully estimating the resultant broken-phase masses to first order in $1/N_c$. 

\section{Acknowledgments}\label{sec:acknowledgments}
L. C. R. W., thanks P. Argyres and G. Semenoff for valuable discussions.  We thank R. Mann for helpful suggestions.  L. S. thanks the Department of Physics at the University of Cincinnati for financial support in the form of the Violet M. Diller Fellowship.  The research of L. C. R. W has been performed in part at the Aspen Center for Physics, which is supported by National Science Foundation Grant  No. PHY-1607611.

\appendix
\section{Analysis of the Tree Potential}\label{sec:app_masses}
We first show that this potential is bounded below when $h \geq 0$ and $f+h \geq 0$, and then discuss symmetry breaking with the further restriction $h > 0$. Next, employing the condition $f+h = 0$, which can be attained naturally through RG flow when $x_s > \bar{x}_s$, we determine the masses of the scalars and gauge bosons. We suppress the RG scale-dependence of the couplings here.

The potential is given by,
\begin{align}\label{eq:class_pot}
V_0 = \tilde{h} \text{Tr} \left(\phi^\dagger \phi \phi^\dagger \phi \right) + \tilde{f} \left[ \text{Tr} \left(\phi^\dagger \phi\right)\right]^2
\end{align}
where $\tilde{h}$ and $\tilde{f}$ are related to $h$ and $f$ by Eq. (\ref{eq:scalings}) and $\phi$ is an $N_c \times N_s$ matrix of complex valued fields,
\begin{align}\label{eq:class_phi}
\phi = 
\begin{pmatrix}
P
\\
Q
\end{pmatrix}.
\end{align}
Here, $P$ is an $N_s \times N_s$ complex valued matrix and $Q$ is an $\left(N_c - N_s\right) \times N_s$ complex valued matrix.  The potential then becomes,
\begin{align}\label{eq:class_pot_1}
V_0 &= \tilde{h} \text{Tr}\left(P^\dagger P P^\dagger P + P^\dagger P Q^\dagger Q + Q^\dagger Q P^\dagger P + Q^\dagger Q Q^\dagger Q \right) 
\nonumber \\
&\quad+ \tilde{f} \left[\text{Tr}\left(P^\dagger P + Q^\dagger Q\right)\right]^2
\end{align}

\subsection{Stability}\label{sec:app_stab}
Using the singular decomposition theorem \cite{Golub_etal}, $\phi$ (taken to be a space-time constant in computing the potential) can be written as $\phi = U^{\dagger}\Sigma V$where $U$ and $V$ are unitary matrices and $\Sigma$ is an $N_c \times N_s$ matrix given by
\begin{align}
\Sigma =
\begin{pmatrix}
\rho_1 & \cdots & 0
\\
\vdots & \ddots & \vdots
\\
0 & \cdots & \rho_{N_s}
\\
\cdots & 0 & \cdots
\\
\vdots & & \vdots
\\
\cdots & 0 & \cdots
\end{pmatrix}.
\end{align}
The $\rho_i$ entries are constants which, without loss of generality, can be taken to be real and positive semidefinite. The potential then becomes,
\begin{align}
V_0 = \tilde{h} \sum_{i=1}^{N_s}\rho_i^4 + \tilde{f}\left(\sum_{i=1}^{N_s} \rho_i^2\right)^2.
\end{align}
Making a change of variable $x_i = \rho_i^2$ for all $i$, the potential can be written as,
\begin{align}\label{eq:class_pot_2}
V_0 = \langle x| \left(\tilde{h} I + \tilde{f} J \right) |x\rangle,
\end{align}
where $|x\rangle = |\rho_1^2, ..., \rho^2_{N_s}\rangle$, $I$ is the $N_s \times N_s$ identity matrix and $J$ is given by
\begin{align}\label{eq:Jmat}
J = |k \rangle \langle k|, \qquad |k\rangle = \begin{pmatrix}1\\ \vdots \\ 1\end{pmatrix}_{N_s \times 1}.
\end{align}

It is evident that $J |k\rangle = N_s |k\rangle$ while any vector $|y_i\rangle$ with $\langle k|y_i\rangle = 0$ yields $J|y_i\rangle = 0$.  Therefore, there is one eigenvector of $J$ with eigenvalue $N_s$ and $N_s - 1$ eigenvectors of $J$ with eigenvalue zero. The matrix in Eq. (\ref{eq:class_pot_2}), $\tilde{h}I + \tilde{f} J$, then has two eigenvalues
\begin{enumerate}
\item $\tilde{h} = \frac{(4\pi)^2 h}{N_c}$
\item $\tilde{h} + N_s \tilde{f} = \frac{(4\pi)^2}{N_c}\left(h+f\right)$
\end{enumerate}
Therefore, if $h \geq0$ and $h+f\geq0$, the potential is bounded from below.

\subsection{Symmetry breaking}\label{app:symm_break}
To discuss symmetry breaking, we first note that Eq. (\ref{eq:class_pot_2}) can be rewritten as,
\begin{align}
V_0 = \langle x|\left[\tilde{h}\left(\frac{J}{N_s} + \sum_{i=1}^{N_s-1}|y_i\rangle \langle y_i| \right)+\tilde{f}J\right]|x\rangle.
\end{align}
Now interpreting the $\rho_i$ as extremal values that minimize the potential, we define,
\begin{align}
\langle x|k\rangle = \sum_{i=1}^{N_s} \rho_i^2 \equiv N_s v^2,
\end{align}
giving,
\begin{align}
V_0 =(4\pi)^2\left\{\frac{h}{N_c} \sum_{i=1}^{N_s-1}\langle x|y_i\rangle \langle y_i|x\rangle + \left(h+f\right) x_s v^4\right\},
\end{align}
with $v \geq 0$. For $h > 0$, and with fixed values for $f + h$ and
$v^2$, this expression is minimized when all $\langle y_{i}|x\rangle = 0$, that is,
when all $|x\rangle \propto |k\rangle$, meaning that all $\rho_ i$ are equal.
Therefore, the minimum potential configuration is given by Eq. (\ref{eq:class_phi})
with $P = vI$ and $Q = 0$, the color-flavor-locked direction. We build
on this result in Sec. \ref{sec:1loop}, including the full one-loop effective
potential, concluding that if $x_s < \bar{x}_s$ then $v=0$, while if $x_s
 > \bar{x}_s$ then $v>0$ with a value determined by the RG scale
$\Lambda_{IR}$ defined there.

\subsection{Scalar masses}
We analyze quadratic fluctuations around the vacuum configuration $P = vI , Q = 0$. The energy of the vacuum configuration is $(h+f) x_{s}(4\pi)^{2}v^4$, vanishing when $h+f = 0$. The mixed terms $\text{Tr}(P^{\dagger}P Q^{\dagger} Q)$ and $(\text{Tr}P^{\dagger}P)\times (\text{Tr} Q^{\dagger} Q)$, do not mix quadratic vacuum fluctuations of the fields in the $P$ and $Q$ sectors.  Therefore, we consider fluctuations in two separate cases:
\begin{enumerate}
\item $P = v I$ with $Q \neq 0$
\item $P = v I + \tilde{P}$ with $Q = 0$.
\end{enumerate}

\subsubsection{Case 1:  $P = v I$ with $Q \neq 0$}
In this case, the field is,
\begin{align}\label{eq:class_case_1}
\phi =
\begin{pmatrix}
v I
\\
Q
\end{pmatrix},
\end{align}
where $v > 0$ is a real value and $I$ is the $N_s \times N_s$ identity matrix.  There is a total of $2 N_c N_s - 2 N_s^2$ degrees of freedom to consider.  Using Eqs. (\ref{eq:class_pot_1}) and (\ref{eq:class_case_1}) we see that all terms in the potential up through quadratic order in $Q$ vanish when $f+h=0$.  The $Q$ fluctuations are massless. 

\subsubsection{Case 2:  $P = vI + \tilde{P}$ with $Q = 0$}
We next consider the case,
\begin{align}
\phi = 
\begin{pmatrix}
vI + \tilde{P}
\\
0
\end{pmatrix},
\end{align}
where
\begin{align}
\tilde{P} = Z + A
\end{align}
with
\begin{align}\label{eq:class_Z}
Z &= 
\begin{pmatrix}
Z_1 & \cdots & 0
\\
\vdots & \ddots & \vdots
\\
0 & \cdots & Z_{N_s}
\end{pmatrix}
\nonumber \\
A &=
\begin{pmatrix}
0 & A_{12} & \cdots & A_{1 N_c}
\\
\vdots & \vdots & \ddots & \vdots
\\
A_{N_c 1} & \cdots & A_{N_c N_c-1} & 0
\end{pmatrix}
\end{align}

The $\tilde{f}$ term of the potential (Eq. (\ref{eq:class_pot})) is given by,
\begin{align}
\tilde{f}\left[\text{Tr}\left(\phi^\dagger \phi\right)\right]^2 &= \tilde{f} \Big\{N_s v^2 + v \text{Tr}\left(Z + Z^\dagger \right) 
\nonumber \\
&\quad+ \text{Tr}\left(Z^\dagger Z + A^\dagger A\right)\Big\}^2.
\end{align}
Since the mixed terms between $Z$ and $A$ are cubic or higher, there are no mixed-mass terms arising here.  The $\tilde{h}$ term of the potential (Eq. (\ref{eq:class_pot})) is given by,
\begin{widetext}
\begin{align}
\tilde{h} \text{Tr}\left(\phi^\dagger \phi \phi^\dagger \phi\right) = \tilde{h}\text{Tr}\left\{\left[v^2 I + v\left(Z + Z^\dagger + A + A^\dagger\right) + Z^\dagger Z + A^\dagger A + Z^\dagger A + A^\dagger Z\right]^2\right\}.
\end{align}
\end{widetext}
The mixed terms quadratic in the fields vanish so we can treat $Z$ and $A$ fluctuations separately when calculating the mass terms.

Therefore, we again separate this case into two further cases:
\begin{enumerate}[(a)]
\item $Z\neq 0$ with $A = 0$
\item $Z = 0$ with $A \neq 0$.
\end{enumerate}

\noindent\textbf{Case 2a:  $Z\neq 0$ with $A = 0$}
The field is,
\begin{align}
\phi =
\begin{pmatrix}
v I + Z
\\
0
\end{pmatrix}.
\end{align}
The terms of the potential that are quadratic in the fields are,
\begin{align}\label{eq:class_2a_pot}
V_{0,\text{quad}} &= (4\pi v)^2 \frac{h}{N_c N_s} \Big\{N_s \text{Tr} \left[\left(Z+Z^\dagger\right)^2\right] 
\nonumber \\
&\quad- \left[\text{Tr}\left(Z+Z^\dagger\right)\right]^2 \Big\}.
\end{align}

Now, taking $Z$ to be given by Eq. (\ref{eq:class_Z}) with $Z_i = X_i + i \, Y_i$, the quadratic terms of the potential (Eq. (\ref{eq:class_2a_pot})) become,
\begin{align}
V_{0,\text{quad}} =(4\pi v)^2 \frac{4 h}{N_c N_s} \langle X| \left(N_s I - J\right) | X \rangle
\end{align}
where $|X\rangle = |X_1,...,X_{N_s}\rangle$, $I$ is the $N_s \times N_s$ identity matrix and $J$ is given by Eq. (\ref{eq:Jmat}).  Notice that the imaginary components of $Z$ (namely $Y_1, ..., Y_{N_s}$) have completely disappeared, indicating that they are massless.

Recall from Sec. \ref{sec:app_stab} that $J$ has one eigenvector (namely the dilaton) with eigenvalue $N_s$ and $N_s - 1$ eigenvectors with eigenvalue zero.  Our mass matrix is proportional to $N_s I_{N_s \times N_s} - J$.  In this case, there is one eigenvector of the mass matrix with eigenvalue zero (namely $|k\rangle$) and $N_s - 1$ eigenvectors with eigenvalue $N_s$.  The vector $|k\rangle$ corresponds to the dilaton, which is massless at this level but, as discussed in Sec. \ref{sec:symm}, gains a mass at the one-loop level.  Therefore, there is a total of $N_s+1$ massless modes from diagonal fluctuations.

The Lagrangian is then (only taking into account quadratic terms of the potential),
\begin{align}
\mathcal{L} = \sum_i\left[\frac{1}{2}\partial_\mu \sigma_i \partial^\mu \sigma_i + \frac{1}{2}\partial_\mu \omega_i \partial^\mu \omega_i + \left(4\pi v\right)^2 \frac{4 h }{N_c} \frac{1}{2}\sigma_i^2 \right],
\end{align}
where we have defined $X_i = 1/\sqrt{2}\sigma_i$ and $Y_i = 1/\sqrt{2}\omega_i$.  The massive modes all have squared masses equal to $4 h (4 \pi v)^2/N_c$.  

\noindent\textbf{Case 2b:  $Z = 0$ with $A \neq 0$}

The field $\phi$ is now,
\begin{align}
\Phi = 
\begin{pmatrix}
vI + A
\\
0
\end{pmatrix},
\end{align}
where $A$ is given by Eq. (\ref{eq:class_Z}).  The terms of the potential that are quadratic in the fields are,
\begin{align}
V_{0,\text{quad}} &= (4\pi v)^2 \frac{h}{N_c}\text{Tr}\left[\left(A+A^\dagger\right)^2\right]
\nonumber \\
&= (4\pi v)^2\frac{h}{N_c}\sum_{ij} \left(A_{ij} + \bar{A}_{ji}\right)\left(A_{ji} + \bar{A}_{ij}\right)
\nonumber\\
&= (4\pi v)^2 \frac{2 h}{N_c}\sum_{i<j}\left|A_{ij} + \bar{A}_{ji}\right|^2.
\end{align}
Therefore, we end up with $N_s^2 - N_s$ pairs of coefficients that are massive, and $N_s^2 - N_s$ that are massless.  

The Lagrangian is then (again taking only quadratic terms in the potential),
\begin{align}
\mathcal{L} &=\sum_{ij}\Big(\partial_\mu X_{ij}\partial^\mu X_{ij} + \partial_\mu X_{ji}\partial^\mu X_{ji} 
\nonumber \\
&\quad+ \partial_\mu Y_{ij}\partial^\mu Y_{ij} + \partial_\mu Y_{ji}\partial^\mu Y_{ij} \Big)
\nonumber \\
&\quad+ (4\pi v)^2 \frac{2 h}{N_c}\sum_{i<j} \left[\left( X_{ij}+X_{ji}\right)^2 + \left(Y_{ij} - Y_{ji}\right)^2\right],
\end{align}
where we have defined $A_{ij} = X_{ij} + i X_{ij}$.  Defining $\alpha_{ij} = \left(X_{ij} + X_{ji}\right)$, $\sigma_{ij} = \left(X_{ij} - X_{ji}\right)$, $\omega_{ij} = \left(Y_{ij} + Y_{ji}\right)$, and $\beta_{ij}=\left(Y_{ij} - Y_{ji}\right)$, the Lagrangian becomes,
\begin{align}
\mathcal{L} &=\sum_{ij}\Big(\frac{1}{2}\partial_\mu \alpha_{ij} \partial^\mu \alpha_{ij} + \frac{1}{2}\partial_\mu \sigma_{ij}\partial^\mu \sigma_{ij} 
\nonumber \\
&\quad+ \frac{1}{2}\partial_\mu \omega_{ij}\partial^\mu \omega_{ij} + \frac{1}{2}\partial_\mu \beta_{ij}\partial^\mu  \beta_{ij} \Big)
\nonumber \\
&\quad+ (4\pi v)^2 \frac{4 h}{N_c}\sum_{ij} \left[ \frac{1}{2}\alpha_{ij}^2 + \frac{1}{2}\beta_{ij}^2\right].
\end{align} 
The massive modes all have squared masses equal to $4 h (4\pi v)^2/N_c$.

\subsubsection{Summary}
\begin{enumerate}
\item Case 1:  $P = v I$ with $Q \neq 0$: $2 N_c N_s - 2 N_s^2$ massless modes; 0 massive modes
\item Case 2(a):  $P = vI + Z$ with $Q = 0$: $N_s+1$ massless modes; $N_s - 1$ massive modes with $m^2 = 4 h (4 \pi v)^2 /N_c$
\item Case 2(b):  $P = vI + A$ with $Q = 0$: $N_s^2 - N_s$ massless modes; $N_s^2 - N_s$ massive modes with $m^2 = 4 h (4 \pi v)^2/N_c$
\end{enumerate}
This leaves a total of $2 N_c N_s - N_s^2 +1$ massless scalar degrees of freedom.  The $N_s^2 - 1$ massive scalars are in the adjoint representation of the unbroken $SU(N_s)$ global symmetry.

\subsection{Gauge-boson masses}
We evaluate the coupling between the vacuum configuration (Eq. (\ref{eq:class_phi}) with $P = v I$ and $Q = 0$) and the $SU(N_c)$ gauge fields $A_\mu^a$ with $a = 1,...,N_c^2-1$.  With $D_\mu = \partial_\mu - i g A_\mu^a t^a$, the coupling term is given by,
\begin{align}
\mathcal{L}_A &= g^2 \sum_{a,b} \text{Tr}\left[\left(t^a \phi\right)^\dagger t^b \phi A_\mu^a A^{\mu,b}\right] 
\nonumber \\
&= g^2 v^2 \sum_{a,b} \text{Tr}\left[K t^{a \dagger} t^b A_\mu^a A^{\mu,b}\right],
\end{align}
where $K$ is an $N_c \times N_s$ matrix with the $N_s \times N_s$ identity matrix in the upper left block and zero elsewhere.  The term $\text{Tr}\left(K t^a t^b\right)$ is zero when $a\neq b$ except when $t^a$ and $t^b$ are certain distinct pairs of generators whose nonzero elements are in identical positions of the off-diagonal $N_s \times (N_c-N_s)$ and $(N_c-N_s)\times N_s$ blocks.  For such generators, $\text{Tr}\left( K t^a t^b\right) + \text{Tr}\left( K t^bt^a\right) = 0$ and their contribution to the sum cancels.

Therefore, the coupling is diagonal in $a$ and $b$.  Evaluating the traces and summing over $a$, we obtain,
\begin{align}
\mathcal{L}_A = g^2 v^2 \sum_{a=1}^{N_c^2-1} f(a) A_a^2,
\end{align}
where $f(a) = 1/2$ if $t^a$ has nonzero entries only in the upper $N_s \times N_s$ diagonal block and $f(a) = 1/4$ if $t^a$ has nonzero entries in the off-diagonal $N_s \times(N_c - N_s)$ blocks.  Finally, $f(a) = (N_c - N_s)/(2 N_c)$ if $t^a$ has nonzero entries in both the upper and lower diagonal blocks and $f(a) = 0$ if $t^a$ has nonzero entries only in the lower $(N_c - N_s) \times (N_c-N_s)$ diagonal block.  

Thus, there are $N_s^2 - 1$ gauge bosons with mass $\lambda (4\pi v)^2/N_c$, $2 N_s (N_c - N_s)$ gauge bosons with mass $\lambda (4\pi v)^2/(2N_c)$, and one gauge boson with mass $\lambda (4\pi v)^2 (N_c - N_s)/(2N_c^2)$.

\bibliography{walking}

\providecommand{\href}[2]{#2}\begingroup\raggedright\begin{thebibliography}{10}

\bibitem{Benini:2019dfy}
F.~Benini, C.~Iossa and M.~Serone, \emph{{Conformality Loss, Walking, and 4D
  Complex Conformal Field Theories at Weak Coupling}},
  \href{https://doi.org/10.1103/PhysRevLett.124.051602}{\emph{Phys. Rev. Lett.}
  {\bfseries 124} (2020) 051602}
  [\href{https://arxiv.org/abs/1908.04325}{{\ttfamily 1908.04325}}].

\bibitem{Hansen:2017pwe}
F.~F. Hansen, T.~Janowski, K.~Lang\ae{}ble, R.~B. Mann, F.~Sannino, T.~G.
  Steele et~al., \emph{{Phase structure of complete asymptotically free
  SU($N_c$) theories with quarks and scalar quarks}},
  \href{https://doi.org/10.1103/PhysRevD.97.065014}{\emph{Phys. Rev. D}
  {\bfseries 97} (2018) 065014}
  [\href{https://arxiv.org/abs/1706.06402}{{\ttfamily 1706.06402}}].

\bibitem{Kaplan:2010zz}
D.~B. Kaplan, J.~W. Lee, D.~T. Son and M.~A. Stephanov, \emph{{Conformality
  lost}}, \href{https://doi.org/10.1142/S0217751X1004872X}{\emph{Int. J. Mod.
  Phys. A} {\bfseries 25} (2010) 422}.

\bibitem{Sannino:2016sfx}
F.~Sannino, A.~Strumia, A.~Tesi and E.~Vigiani, \emph{{Fundamental partial
  compositeness}}, \href{https://doi.org/10.1007/JHEP11(2016)029}{\emph{JHEP}
  {\bfseries 11} (2016) 029}
  [\href{https://arxiv.org/abs/1607.01659}{{\ttfamily 1607.01659}}].

\bibitem{Machacek:1984zw}
M.~E. Machacek and M.~T. Vaughn, \emph{{Two Loop Renormalization Group
  Equations in a General Quantum Field Theory. 3. Scalar Quartic Couplings}},
  \href{https://doi.org/10.1016/0550-3213(85)90040-9}{\emph{Nucl. Phys. B}
  {\bfseries 249} (1985) 70}.

\bibitem{Gardi:1998qr}
E.~Gardi, G.~Grunberg and M.~Karliner, \emph{{Can the QCD running coupling have
  a causal analyticity structure?}},
  \href{https://doi.org/10.1088/1126-6708/1998/07/007}{\emph{JHEP} {\bfseries
  07} (1998) 007} [\href{https://arxiv.org/abs/hep-ph/9806462}{{\ttfamily
  hep-ph/9806462}}].

\bibitem{Corless_etal}
R.~M. Corless, G.~H. Gonnet, D.~E.~G. Hare and et~al., \emph{{On the LambertW
  function}}, \href{https://doi.org/10.1007/BF02124750}{\emph{Adv. Comput.
  Math} {\bfseries 5} (1996) 329}.

\bibitem{Pomoni:2009joh}
E.~Pomoni and L.~Rastelli, \emph{{Large N Field Theory and AdS Tachyons}},
  \href{https://doi.org/10.1088/1126-6708/2009/04/020}{\emph{JHEP} {\bfseries
  04} (2009) 020} [\href{https://arxiv.org/abs/0805.2261}{{\ttfamily
  0805.2261}}].

\bibitem{Grinstein:2011dq}
B.~Grinstein and P.~Uttayarat, \emph{{A Very Light Dilaton}},
  \href{https://doi.org/10.1007/JHEP07(2011)038}{\emph{JHEP} {\bfseries 07}
  (2011) 038} [\href{https://arxiv.org/abs/1105.2370}{{\ttfamily 1105.2370}}].

\bibitem{Antipin:2011aa}
O.~Antipin, M.~Mojaza and F.~Sannino, \emph{{Light Dilaton at Fixed Points and
  Ultra Light Scale Super Yang Mills}},
  \href{https://doi.org/10.1016/j.physletb.2012.04.050}{\emph{Phys. Lett. B}
  {\bfseries 712} (2012) 119}
  [\href{https://arxiv.org/abs/1107.2932}{{\ttfamily 1107.2932}}].

\bibitem{Antipin:2012sm}
O.~Antipin, M.~Mojaza and F.~Sannino, \emph{{Jumping out of the light-Higgs
  conformal window}},
  \href{https://doi.org/10.1103/PhysRevD.87.096005}{\emph{Phys. Rev. D}
  {\bfseries 87} (2013) 096005}
  [\href{https://arxiv.org/abs/1208.0987}{{\ttfamily 1208.0987}}].

\bibitem{Gildener:1975cj}
E.~Gildener, \emph{{Radiatively Induced Spontaneous Symmetry Breaking for
  Asymptotically Free Gauge Theories}},
  \href{https://doi.org/10.1103/PhysRevD.13.1025}{\emph{Phys. Rev. D}
  {\bfseries 13} (1976) 1025}.

\bibitem{Coleman:1973jx}
S.~R. Coleman and E.~J. Weinberg, \emph{{Radiative Corrections as the Origin of
  Spontaneous Symmetry Breaking}},
  \href{https://doi.org/10.1103/PhysRevD.7.1888}{\emph{Phys. Rev. D} {\bfseries
  7} (1973) 1888}.

\bibitem{Yamagishi:1981qq}
H.~Yamagishi, \emph{{Coupling Constant Flows and Dynamical Symmetry Breaking}},
  \href{https://doi.org/10.1103/PhysRevD.23.1880}{\emph{Phys. Rev. D}
  {\bfseries 23} (1981) 1880}.

\bibitem{Gildener:1976ih}
E.~Gildener and S.~Weinberg, \emph{{Symmetry Breaking and Scalar Bosons}},
  \href{https://doi.org/10.1103/PhysRevD.13.3333}{\emph{Phys. Rev. D}
  {\bfseries 13} (1976) 3333}.

\bibitem{Hill:2014mqa}
C.~T. Hill, \emph{{Is the Higgs Boson Associated with Coleman-Weinberg
  Dynamical Symmetry Breaking?}},
  \href{https://doi.org/10.1103/PhysRevD.89.073003}{\emph{Phys. Rev. D}
  {\bfseries 89} (2014) 073003}
  [\href{https://arxiv.org/abs/1401.4185}{{\ttfamily 1401.4185}}].

\bibitem{Golub_etal}
G.~H. Golub and C.~F. Van~Loan, \emph{Matrix Computations, 3rd ed.} Johns
  Hopkins University Press, Baltimore, MD, 1996.

\end{thebibliography}\endgroup
\bibliographystyle{JHEP}

\end{document}